\documentclass[preprint,12pt,authoryear]{elsarticle}
\usepackage[utf8]{inputenc}
\usepackage{xcolor}

\usepackage{amssymb}
\usepackage{graphicx}
\usepackage{epstopdf, epsfig}

\usepackage{hyperref}
\usepackage[english]{babel}

\newcommand{\RomanNumeralCaps}[1]
\linenumbers
\usepackage{amsmath}
\usepackage{amssymb}
\usepackage{natbib}

\AtBeginDocument{}
\AtBeginDocument{}
\AtBeginDocument{}

\journal{European Journal of Mechanics - B/Fluids}

\title{The Objective Deformation Component \\ of a Velocity Field}

\author[inst1]{Bálint Kaszás}
\affiliation[inst1]{organization={Institute for Mechanical Systems, ETH Zurich},
            city={Zurich},
            country={Switzerland}}
            
\author[inst2]{Tiemo Pedergnana}
\affiliation[inst2]{organization={Institute for Energy and Process Engineering, ETH Zurich},
            city={Zurich},
            country={Switzerland}}
\author[inst1]{George Haller}
\begin{document}

\begin{abstract}
For an arbitrary velocity field \textbf{$\mathbf{v}$ }defined on
{a finite, fixed spatial domain}, we find the closest rigid-body velocity field\textbf{
$\mathbf{v}_{RB}$ }to $\mathbf{v}$ in the $L^2$ norm.
The resulting deformation velocity component, $\mathbf{v}_{d}=\mathbf{v-\mathbf{v}}_{RB}$,
turns out to be { frame-indifferent} and physically observable.
Specifically, if $\mathbf{Q}_{\text{RB}}(t)$ is the rotation tensor describing the motion of the closest rigid body frame, then $\mathbf{v}$
is seen as $\mathbf{Q}_{\text{RB}}^{T}\mathbf{v}_{d}$ by
an observer in that frame. As a consequence, the momentum, energy,
vorticity, enstrophy, and helicity of the flow all become { frame-indifferent}
when computed from the deformation velocity component $\mathbf{v}_{d}$. 

\end{abstract}

\maketitle

\section{Introduction\label{sec:Introduction}}

The analysis of fluid flows often starts with the inspection of the
instantaneous spatial features of the velocity field, such as the
streamlines, streamsurfaces, as well as the distribution of vorticity,
rate of strain, enstrophy and various other scalars used in classic
vortex criteria (\cite{perry_topology_1994,haller_objective_2005,haller_defining_2016}).
Virtually all these spatial features, however, depend on the observer,
and hence do not reflect purely intrinsic properties of the fluid. { The independence of material response on the observer, which is usually referred to as frame-indifference or objectivity,  is a fundamental axiom of continuum mechanics (\cite{Gurtin1981})}.

Based on physical considerations, one may nevertheless argue for a
distinguished frame of reference in which to evaluate customary flow
diagnostics. If such a frame exists for the fluid, it is often called
the co-moving or proper frame (\cite{landau_classical_1975}). 

Finding proper frames is straightforward when there is a homogeneous
(periodic or infinite) direction of flow propagation. This is the
case for traveling waves or relative periodic orbits in shear flows
(\cite{waleffe_exact_2001}) for which a well-defined base velocity
field exists. In some cases, even time-dependent phase speeds $c$
can be found for fluids in a channel, minimizing the difference
between the velocity field at $x$ and at $x-ct$, as in \cite{mellibovsky_travelling_2012, kreilos_comoving_2014}. 

For general fluid flows, finding a single co-moving frame in which
the fluid velocity field $\mathbf{v}(\mathbf{x},t)$ becomes simple
(or even steady) is unrealistic, as pointed out by \cite{lugt_dilemma_1979}. One may
nevertheless seek a rigid-body frame that is overall as close as possible
to being a co-moving frame for the whole fluid. Subtracting such a
rigid-body velocity field $\mathbf{v}_{RB}(\mathbf{\mathbf{x}},t)$
from $\mathbf{v}(\mathbf{x},t)$ would then yield a deformation velocity
field that describes the overall deviation of the fluid from rigid
body motion as closely as possible. In order for such a decomposition
to be meaningful, we propose the following requirements:
\begin{itemize}
\item [{(i)}] $\mathbf{\mathbf{v}}_{RB}$ should be a rigid body velocity
field that is closest to $\mathbf{v}$ in a physically relevant norm.
\item [{(ii)}] $\mathbf{v-\mathbf{v}}_{RB}$ should be frame-indifferent (objective) in order to capture intrinsic features of the fluid (see \cite{truesdell_non-linear_2004}),
\item [{(iii)}] $\mathbf{v-\mathbf{v}}_{RB}$ should be {physically} observable: an
appropriately chosen, single observer of the fluid flow should be
able to measure this deformation velocity field over the whole domain. {Specifically, $\mathbf{v-\mathbf{v}}_{RB}$ should be related to the original velocity field $\mathbf{v}$ under a proper change of observers, i. e. a time dependent Euclidean transformation of the spatial domain.}
\end{itemize}

Available decompositions for velocity fields fail to satisfy these
three requirements. An example is the classic Reynolds decomposition
(see, e.g., \cite{pope_turbulent_2000}), in which velocity field of a turbulent
fluid is written as a sum of its time average and a fluctuating part.
As shown in \cite{speziale_review_1998}, the fluctuating part of
the velocity is objective but not physically observable, {that is, the total velocity and the fluctuating part are not related to each other via a time dependent Euclidean transformation}. The existence
of a well-defined time-average is not guaranteed either. Another example,
the Helmholtz--Hodge (\cite{arfken_mathematical_2005}) decomposition,
splits $\mathbf{v}(\mathbf{x},t)$ into an incompressible and an irrotational
component in a given frame. As a generalization, the Weber--Clebsch
representation (\cite{lamb_hydrodynamics_1945,constantin_eulerian-lagrangian_2001})
has a linear combination of non-potential components. This, however,
returns the same flow when the flow is incompressible without extracting
any of its deformational features. 

The separation of rotations and internal motions is also an important step in the analysis of the $n$-body problem, as discussed, for example, in \cite{Littlejohn97}. There, an appropriate choice of the reference frame makes it possible to write the Lagrangian as a sum of rotational and deformational components.  Further local decompositions seek
to isolate rotational velocity components based on the velocity gradient.
{ Examples include the deformation-rotation decomposition of \cite{Batchelor2000} and  the procedures of \cite{kolar_vortex_2007}, \cite{liu_rortexnew_2018}, \cite{wang_explicit_2019} and \cite{holmedal_spin_2020}}.
In the Lagrangian frame, only infinitesimal decompositions of the
flow have been obtained. The closest of these in spirit to the present
study is the polar decomposition, which yields a polar rotation tensor
that is pointwise the closest rigid-body rotation to the deformation
gradient (\cite{neff_griolis_2014}).

In a recent stream of papers in the scientific visualization community,
(\cite{bujack_topology-inspired_2016,gunther_generic_2017,kim_robust_2019,rojo_vector_2020,gunther_hyper-objective_2020})
seek an objectively defined minimally unsteady component $\mathbf{v}_{*}(\mathbf{x},t)$
of $\mathbf{v}(\mathbf{x},t)$. \cite{haller_can_2021} shows that
the proposed implementations of these principles leads to $\mathbf{v}_{*}(\mathbf{x},t)\equiv\mathbf{v}(\mathbf{x},t)$
and hence fail to provide the desired decomposition. Using a simple
counterexample, \cite{haller_can_2021} also shows that even under
a correct implementation, $\mathbf{v}_{*}(\mathbf{x},t)$ would not
be objective, irrespective of the measure of unsteadiness chosen.
\cite{theisel_vortex_2021} continue to claim objectivity of $\mathbf{v}_{*}(\mathbf{x},t)$,
but their argument uses the pull-back, as opposed to the inverse,
of the nonlinear transformation $\mathbf{v}\mapsto\mathbf{v}_{*}$
when computing $\mathbf{v}_{*}\mapsto\mathbf{v}$. This is incorrect,
given that velocity fields are non-objective precisely because they
do not transform under the push-forward and pull-back of frame changes.\textbf{
}

In a parallel development, \cite{hadwiger_time-dependent_2019}, \cite{rautek_objective_2021}, and \cite{zhang_interactive_2021}
seek a nonlinear (i.e., non-rigid-body) observer velocity field $\mathbf{u}(\mathbf{x},t)$
that is simultaneously close to \textbf{$\mathbf{v}$}, has small
rate of strain and yields a small Lie-derivative for $\mathbf{v}-\mathbf{u}$
along $\mathbf{u}$. While these individual objectives all come with
their own user-dependent weight functions, the relative velocity $\mathbf{v}-\mathbf{u}$
can indeed be shown to be objective, as long as the optimization principle
has a globally unique solution (as assumed but not verified by the
authors). { The frame change represented by $\mathbf{u}$, however,
is nonlinear and hence $\mathbf{v}-\mathbf{u}$ cannot be
observed by any single physical observer. Thus, requirement (iii) is not satisfied.}

An alternative idea is to perform pointwise local velocity field decompositions
in the flow by passing to local frames co-rotating with the eigenvectors
of the rate-of-strain tensor (\cite{astarita_objective_1979,tabor_stretching_1994,lapeyre_does_1999}).
While these pointwise frame changes are individually objective, they
are not observable by a single physical observable and cannot be stitched
up to form a smoothly varying global coordinate change (\cite{haller_can_2021}).
In contrast, the vortex criteria developed in \cite{liu_third_2019, liu_objective_2019-1,liu_objective_2019, liubook}
can be viewed as evaluations of classic vortex criteria in a frame
generated by the spin-deviation tensor, as shown by \cite{haller_can_2021}.
This frame, however, is not obtained as a closest rigid body frame
from any systematic optimization and hence fails to satisfy the requirement
(i) above. 

In this paper, we derive a closed form solution for a rigid-body velocity
component $\mathbf{v}_{RB}(\mathbf{\mathbf{x}},t)$ satisfying the
requirements (i)-(iii) on a bounded spatial domain $U\subset\mathbb{R}^{3}$
. We achieve this by explicitly solving the underlying optimization
problem defining $\mathbf{v}_{RB}(\mathbf{\mathbf{x}},t)$. We then
show that $\mathbf{v}_{d}=\mathbf{v-\mathbf{v}}_{RB}$ is observer-indifferent
and directly observable in a specific Euclidean observer frame. The rigid body velocity
$\mathbf{v}_{RB}$  is closest to $\mathbf{v}$ in the classic $L^{2}$
norm.

All classic non-objective vortex criteria become objective when evaluated
in the frame co-moving with $\mathbf{v}_{RB}(\mathbf{\mathbf{x}},t)$. { However, it is still not guaranteed that the objectivized vortex-criteria correctly identify all eddies. Specifically, while after objectivization all observers will agree on the conclusions of the vortex criteria, the classic criteria we have reviewed still lack a rigorous connection to material behavior in the fluid. }

In addition, passage to the deformation velocity preserves the rate-of-strain
tensor of the full velocity field, causing $\textbf{\ensuremath{\mathbf{v}_{d}}}$
and $\mathbf{v}$ to have the same objective Eulerian coherent structures
(OECSs), as defined by \cite{serra_objective_2016}. Similarly, the
instantaneous vorticity deviation (IVD) defined by \cite{haller_defining_2016},
another objective indicator of Eulerian coherence, is also preserved
from $\mathbf{v}$ for $\textbf{\ensuremath{\mathbf{v}_{d}}}$
by our decomposition. 

Perhaps most importantly, all common non-objective Eulerian scalar
quantities, such as the kinetic energy, enstrophy, and helicity, become
observer-independent when they are evaluated on the deformation velocity
$\textbf{\ensuremath{\mathbf{v}_{d}}}(\mathbf{\mathbf{x}},t)$.
This gives a natural way to define the deformation kinetic energy,
deformation enstrophy and deformation helicity of the flow as intrinsic
physical quantities that do not depend on the observer.

{ The optimization with respect to the physically motivated $L^2$ norm yields a rigid-body velocity field, $\mathbf{v}_{RB}$, that is uniquely determined over a given spatial domain $U$. For most problems, the velocity field is a priori given over a certain domain of interest, which defines $U$ as well. Therefore the deformation velocity $\mathbf{v}_d = \mathbf{v}-\mathbf{v}_{RB}$ can be uniquely associated to the velocity field $\mathbf{v}$. 

However, when one has some freedom in choosing the spatial domain $U$, we show that this choice can influence the deformation velocity $\mathbf{v}_d$. Specifically, if one chooses a large enough domain for a complicated flow, it is unlikely that passing to a single distinguished reference frame could eliminate all rotating motions simultaneously. {Through an example, we show that in oceanographic applications this is indeed the case. Due to the multitude of mesoscale eddies, there is no optimal way to isolate a single rigid-body observer and we obtain $\mathbf{v}_{RB}\approx 0$. If, however, smaller domains with well defined rotational features are of interest in a given problem, then the method successfully extracts a non-zero rigid-body velocity. Choosing the spatial domain $U$ optimally remains a challenge for general flows. }

\section{Main results \label{sec:Main-results}}

We seek the rigid body velocity field $\mathbf{v}_{RB}(\mathbf{x},t)$
closest to an arbitrary mass-preserving velocity field $\mathbf{v}(\mathbf{x},t)$
defined on a bounded (and potentially time-dependent) spatial domain
$U\subset\mathbb{R}^{3}$. To measure closeness between two velocity
fields, we use the $L^{2}$ norm 

\begin{equation}
\left\Vert \mathbf{f}\right\Vert _{L^{2}}^{2}:=\frac{1}{M}\int_{U}\left|\mathbf{f}(\mathbf{x},t)\right|^{2}dm,\label{eq:sobolevnorm}
\end{equation}
where $M$ is the total mass contained in the domain $U$. { We use a mass-based, as
opposed to the customary volume-based, norm to cover compressible but mass-preserving 
flows. For incompressible flows, our mass-based minimization is equivalent to a volume-based minimization. }

As any rigid body velocity field, the $\mathbf{v}_{RB}$ we seek must
have the general form
\begin{equation}
\mathbf{v}_{RB}(\mathbf{\mathbf{x}},t)=\dot{\mathbf{x}}_{A}(t)+\boldsymbol{\Omega}(t)\left(\mathbf{x}-\mathbf{x}_{A}(t)\right)=\dot{\mathbf{x}}_{A}(t)+\boldsymbol{\omega}(t)\times\left(\mathbf{x}-\mathbf{x}_{A}(t)\right),\label{eq:vrb}
\end{equation}
where $\mathbf{x}_{A}(t)\in U$ is the current position of a material
point on the rigid body whose instantaneous velocity is $\dot{\mathbf{x}}_{A}(t)$;
$\mathbf{x\in}U$ is another, arbitrary material point on the rigid
body, whose instantaneous velocity is $\mathbf{v}_{RB}(\mathbf{\mathbf{x}},t)$.
The vector $\boldsymbol{\omega}(t)\in\mathbb{R}^{3}$ denotes the
angular velocity of the rigid body and the skew-symmetric tensor $\boldsymbol{\Omega}(t)=-\boldsymbol{\Omega}^{T}(t)\in\mathbb{R}^{3\times3}$
is defined as 

\begin{equation}
\boldsymbol{\Omega}(t)\mathbf{e}=\boldsymbol{\omega}(t)\times\mathbf{e},\quad\forall\mathbf{e}\in\mathbb{R}^{3}.\label{eq:axvector}
\end{equation}

More generally, a vector $\boldsymbol{\omega}(t)$ defined by the
relation (\ref{eq:axvector}) is called the \emph{dual vector} of a
skew-symmetric matrix $\boldsymbol{\Omega}(t)$ (see, e.g., \cite{arfken_mathematical_2005}).
We will use $\boldsymbol{\omega}(t)$ and $\boldsymbol{\Omega}(t)$
interchangeably: all results formulated in terms of the tensor $\boldsymbol{\Omega}(t)$
can be recast in terms of $\boldsymbol{\omega}(t)$ using the formula
(\ref{eq:axvector}) and vice versa. 

We will show that the tuple $(\boldsymbol{\Omega}(t),\mathbf{x}_{A}(t))$,
for which the distance
\begin{equation}
L(\boldsymbol{\Omega},\mathbf{\mathbf{x}}_{A},t)=\left\Vert \mathbf{\mathbf{v}}\left(\mathbf{x},t\right)-\mathbf{v}_{RB}\left(\mathbf{x},t\right)\right\Vert _{L^{2}}^{2}\label{eq:functional general def}
\end{equation}
is minimal, can be computed explicitly in terms of only the velocity
field $\mathbf{v}(\mathbf{x},t)$ and its domain of definition $U.$ { To motivate the choice of the $L^2$ norm in \eqref{eq:functional general def}, we note that \eqref{eq:functional general def} is proportional to the kinetic energy of $\mathbf{v}-\mathbf{v}_{RB}$. This means that we seek the rigid-body velocity $\mathbf{v}_{RB}$ that minimizes the kinetic energy of the velocity field seen from that rigid body. More general (albeit less physical) norms could also be considered, but they do not change the results significantly, as we show in Appendix G.}

The final result from the minimization of $L(\boldsymbol{\Omega},\mathbf{\mathbf{x}}_{A},t)$
will depend on the moment of inertia tensor
\begin{equation}
\mathbf{\boldsymbol{\Theta}:=}M\overline{\left(\left|\mathbf{x}-\bar{\mathbf{x}}\right|^{2}\right)\mathbf{I}-(\mathbf{x}-\bar{\mathbf{x}})\otimes(\mathbf{x}-\bar{\mathbf{x}})},\label{eq:moment of inertia}
\end{equation}
where overbar denotes mass-based averaging over $U$ (as in (\ref{eq:sobolevnorm})),
\textbf{$\mathbf{I}\in\mathrm{\mathbb{R}^{3\times3}}$ }denotes the
identity tensor. This is exactly the classic moment
of inertia tensor of a rigid body, computed formally with respect
to the center of mass $\bar{\mathbf{x}}$ of the fluid mass filling
$U$.

We show in Section \ref{sec:Derivation-of-the} that, the closed form
solution for the minimizer of (\ref{eq:functional general def}) is
\begin{align}
\mathbf{x}_{A}(t) & =\bar{\mathbf{x}}(t),\label{eq:optimal solution}\\
\boldsymbol{\omega}(t) & =M\boldsymbol{\Theta}{}^{-1}\overline{(\mathbf{x}-\bar{\mathbf{x}})\times(\mathbf{v}-\overline{\mathbf{v}})}.\nonumber 
\end{align}
As we will point out in \ref{sec:Derivation-of-the}, the tuple in
eq. (\ref{eq:optimal solution}) optimizes the $L^{2}$ distance of
$\mathbf{v}_{RB}(\mathbf{\mathbf{x}},t)$ from $\mathbf{v}\left(\mathbf{x},t\right)$
on average over any finite time interval $\left[t_{0},t_{1}\right]$,
not just at a discrete time $t\in\left[t_{0},t_{1}\right].$ 

By eq. (\ref{eq:optimal solution}), the reference point $\mathbf{x}_{A}(t)$
can be chosen as the center of mass of the fluid, and the optimal
rigid-body angular velocity, $\boldsymbol{\omega}(t)$, is
a linear function of the velocity field $\mathbf{v}(\mathbf{x},t)$.
Note that the vector $\mathbf{\boldsymbol{L}}=M\overline{(\mathbf{x}-\bar{\mathbf{x}})\times(\mathbf{v}-\overline{\mathbf{v}})}$
is the angular momentum of the fluid with respect
to its center of mass (\cite{goldstein_classical_2008}). Therefore, the angular velocity $\boldsymbol{\omega}(t)$ of the
closest rigid body motion obeys the equation $\boldsymbol{\Theta}\boldsymbol{\omega}=\mathbf{\boldsymbol{L}},$
as expected from classical rigid-body mechanics. We emphasize, however, that
we have arrived at this result by interpreting the exact solution
of the underlying variational principle rather than by analogy.

We now define the deformation component of the velocity (or \emph{deformation
velocity}), $\textbf{\ensuremath{\mathbf{v}_{d}}}$, as the difference
of $\mathbf{v}$ from $\mathbf{v}_{RB}$:
\begin{equation}
\textbf{\ensuremath{\mathbf{v}_{d}}}(\mathbf{\mathbf{x}},t):=\mathbf{v}(\mathbf{\mathbf{x}},t)-\mathbf{v}_{RB}(\mathbf{\mathbf{x}},t)=\mathbf{v}(\mathbf{\mathbf{x}},t)-\mathbf{\bar{v}}(t)-\boldsymbol{\omega}(t)\times\left(\mathbf{x-}\bar{\mathbf{x}}(t)\right).\label{eq:definition of deformation velocity}
\end{equation}
By construction, $\mathbf{v}_{RB}$ satisfies the requirement (i)
we have laid down for a meaningful decomposition of $\mathbf{v}$
in Section \ref{sec:Introduction}. In Section \ref{sec:Derivation-of-the} we will
show that additional requirements (ii) and (iii) also hold. Specifically,
while the angular velocity $\boldsymbol{\omega}(t)$ is not
objective, $\textbf{\ensuremath{\mathbf{v}_{d}}}(\mathbf{\mathbf{x}},t)$
is nevertheless an objective vector field and can be observed physically
in a frame co-rotating with $\mathbf{v}_{RB}(\mathbf{\mathbf{x}},t)$. 

Discrete versions of eq. \ref{eq:optimal solution} defining the angular velocity vector also appear in the $n$-body problem (see, e.g., \cite{tachibana86} and \cite{marsden92}), where separating rotational and deformational degrees of freedom is essential. In that setting, passing to a (possibly time-dependent) body frame decomposes the kinetic energy into two terms: one comes from rotations, and the other comes from changes in the shape of the $n$-body system (i.e., deformations). \cite{Eckart35} gives a standard method of choosing the body frame. Moreover, this separation of the kinetic energy (and hence the $n$-body Lagrangian) is shown in \cite{Littlejohn97} to be gauge invariant. Since the gauge convention refers to choosing the body frame, gauge invariance is the same as objectivity in the language of continuum mechanics. Our results, however, are indifferent to the underlying Lagrangian of the system. The decomposition defined by eq. \ref{eq:definition of deformation velocity} can be applied to arbitrary velocity fields,  even to those obtained from measurements.

{Velocity fields with similar names have appeared in other contexts before, such as \cite{Bergeron1928}'s deformation field for frontogenesis (see also \cite{Gurtin1981} and \cite{Pedlosky}). However, in that case, {\em deformation field} refers to a specific velocity field that induces deformation. Our equation \eqref{eq:definition of deformation velocity} instead defines a deformation velocity associated to any given velocity field $\mathbf{v}$. }

The results outlined above enable us to \emph{objectivize} a number
of originally non-objective scalar-, vector- and tensor-fields derived
from $\mathbf{v}(\mathbf{x},t)$ by computing these fields in the
frame co-moving with $\mathbf{v}_{RB}$. In practice, this simply
means computing the fields from $\textbf{\ensuremath{\mathbf{v}_{d}}}(\mathbf{\mathbf{x}},t)$
as opposed to $\mathbf{v}(\mathbf{\mathbf{x}},t)$. Examples of scalar
fields that become objective in this fashion include the kinetic energy,
enstrophy, and helicity. Similarly, when computed in the frame co-moving
with $\mathbf{v}_{RB}$, the vorticity, linear momentum and angular
momentum become objective vector fields and the velocity gradient
and the spin tensor become objective tensor fields. For instance,
the \emph{deformation kinetic energy}

\begin{equation}
E_{d}(t)=\frac{1}{2}\left|\mathbf{v}_{d}(\mathbf{\mathbf{x}},t)\right|^{2}\label{eq:objective energy}
\end{equation}
is a frame-invariant measure of the kinetic energy related to flow
deformation. One may, for instance, plot this scalar field to reveal observer-independent Eulerian features
of the flow that arise from fluid deformation rather than rigid-body
translation and rotation. 

\section{Derivation of the main results\label{sec:Derivation-of-the}}

We now present the derivation of our main results discussed in Section
\ref{sec:Main-results}. 

\subsection{Solution to the optimization problem}

After substitution of the general expression of the rigid body velocity
field (\ref{eq:vrb}) into the distance functional $L(\boldsymbol{\Omega},\mathbf{\mathbf{x}}_{A},t)$
defined in eq. (\ref{eq:functional general def}), we obtain
\begin{align}
L_{1}(\boldsymbol{\Omega},t) & =\left\Vert \mathbf{\mathbf{v}}\left(\mathbf{x},t\right)-\mathbf{v}_{RB}\left(\mathbf{x},t\right)\right\Vert _{L^{2}}^{2}\nonumber \\
 & =\frac{1}{M}\int_{U}\left\{ \left|\mathbf{v}(\mathbf{\mathbf{x}},t)-\dot{\mathbf{x}}_{A}(t)-\boldsymbol{\Omega}(t)\left(\mathbf{x}-\mathbf{x}_{A}(t)\right)\right|^{2}\right\}\rho(\mathbf{x},t)dV \label{eq:functionalSpecific}
\end{align}
where $\rho(\mathbf{x},t)$ is the density of the (potentially compressible)
fluid. 

As we will see, the solution of the minimization problem $L_{1}(\boldsymbol{\Omega},t)$
will yield a unique closest rigid body motion, but any point of that
body can be chosen as the reference point $\mathbf{x}_{A}(t)$. Indeed,
if a given pairing $\left(\mathbf{x}_{A}(t),\boldsymbol{\Omega}(t)\right)$
generates the closest rigid-body motion to $\mathbf{v}(\mathbf{x},t)$,
then all pairs $\left(\mathbf{x}_{B}(t),\boldsymbol{\Omega}(t)\right)$
will qualify as well, as long as 
\[
\dot{\mathbf{x}}_{B}(t)=\dot{\mathbf{x}}_{A}(t)+\boldsymbol{\Omega}(t)\left(\mathbf{x}_{B}(t)-\mathbf{x}_{A}(t)\right)
\]
holds, i.e., $\left(\mathbf{x}_{B}(t),\boldsymbol{\Omega}(t)\right)$
describe the same rigid body motion. The most straightforward choice
for $\mathbf{x}_{A}(t)$ is the center of mass of the fluid, i.e,
\[
\mathbf{x}_{A}(t)=\bar{\mathbf{x}}(t)=\frac{1}{M}\int_{U}\mathbf{x}\,dm=\frac{1}{M}\int_{U}\mathbf{x}\rho(\mathbf{x},t)\,dV.
\]
As the total mass of the fluid is assumed to be constant, we then
have
\begin{align*}
\dot{\mathbf{x}}_{A}(t) & =\frac{D}{Dt}\bar{\mathbf{x}}(t)=\frac{D}{Dt}\frac{1}{M}\int_{U}\mathbf{x}\,dm=\frac{1}{M}\int_{U}\frac{D}{Dt}\mathbf{x}\,dm\\
 & =\frac{1}{M}\int_{U}\mathbf{v}\left(\mathbf{x},t\right)\,dm=\frac{1}{M}\int_{U}\mathbf{v}\left(\mathbf{x},t\right)\rho(\mathbf{x},t)\,dV\\
 & =\bar{\mathbf{v}}(t),
\end{align*}
where $\bar{\mathbf{v}}(t)$ denotes the mass-based (or density-weighted)
average of the velocity field over $U$. For incompressible flows,
this $\bar{\mathbf{v}}$ also agrees with the spatial average of\textbf{
$\mathbf{v}$ }over the domain\textbf{ $U$}. 

Therefore, the closest rigid-body motion to $\mathbf{v}(\mathbf{x},t)$,
without loss of generality, can be sought in the more specific form
\begin{equation}
\mathbf{v}_{RB}(\mathbf{\mathbf{x}},t)=\bar{\mathbf{v}}(t)+\boldsymbol{\Omega}(t)\left(\mathbf{x}-\bar{\mathbf{x}}(t)\right),\label{eq:explicit vrb}
\end{equation}
where $\boldsymbol{\Omega}(t)$ is the global minimizer of the functional
$L_{1}(\boldsymbol{\Omega},t)$ defined in eq. (\ref{eq:functionalSpecific}).
Such a unique global minimizer must solve the associated Euler--Lagrange
equation
\begin{equation}
\frac{\partial}{\partial\boldsymbol{\Omega}}L_{1}(\boldsymbol{\Omega},t)=0.\label{eq:E-L equation}
\end{equation}
This last equation is also the Euler--Lagrange equation associated
with the problem of extremizing the functional $\frac{1}{t_{1}-t_{0}}\int_{t_{0}}^{t_{1}}L_{1}(\boldsymbol{\Omega}(t),t)dt$.
Therefore, eq. (\ref{eq:E-L equation}) provides a solution $\boldsymbol{\Omega}(t)$
that optimizes the $L^{2}$ distance of $\mathbf{v}_{RB}(\mathbf{\mathbf{x}},t)$
from $\mathbf{v}\left(\mathbf{x},t\right)$ on average over any finite
time interval $\left[t_{0},t_{1}\right]$, not just just at a discrete
time $t\in\left[t_{0},t_{1}\right].$ 

To calculate the minimizing $\boldsymbol{\Omega}(t)$ explicitly,
we express $L_{1}(\boldsymbol{\Omega},t)$ in coordinates. We will
only focus on the additive terms of $L_{1}(\boldsymbol{\Omega},t)$
that depend on $\boldsymbol{\Omega}$, collecting them in a functional
$\widetilde{L_{1}}(\boldsymbol{\Omega},t)$ with $\frac{\partial}{\partial\boldsymbol{\Omega}}\left[L_{1}-\widetilde{L_{1}}\right]\equiv\mathbf{0}$.
This $\widetilde{L_{1}}$ can be written as 
\begin{align}
\widetilde{L_{1}}(\boldsymbol{\Omega},t) & =\frac{1}{M}\int_{U}\left\{ -2(v_{i}-\dot{\bar{x}}_{i})\Omega_{ij}\left(x_{j}-\bar{x}_{j}\right)+\Omega_{ij}\left(x_{j}-\bar{x}_{j}\right)\Omega_{ik}\left(x_{k}-\bar{x}_{k}\right)\right\} 
  \rho dV,\label{eq:L1_in_coordinates}
\end{align}
with summation implied over repeated indices. As we show in Appendix A, 
substitution of eq. (\ref{eq:L1_in_coordinates}) into (\ref{eq:E-L equation})
leads to a lengthy expression which can nevertheless be simplified
in invariant form to 
\begin{equation}
\overline{(\mathbf{x}-\bar{\mathbf{x}})\times(\mathbf{v}-\overline{\mathbf{v}})}=\overline{\left(\left|\mathbf{x}-\bar{\mathbf{x}}\right|^{2}\right)\boldsymbol{\omega}-(\mathbf{x}-\bar{\mathbf{x}})\otimes(\mathbf{x}-\bar{\mathbf{x}})\boldsymbol{\omega}}\label{eq:omega equation}
\end{equation}
where the angular velocity $\boldsymbol{\omega}(t)$ is an
extremizer of $\widetilde{L_{1}}$.
We can write the right-hand side of eq. (\ref{eq:omega equation})
as $\frac{1}{M}\boldsymbol{\Theta}\boldsymbol{\omega}$
using the moment of inertia tensor defined in (\ref{eq:moment of inertia}).
As the classic moment of inertia tensor, $\boldsymbol{\Theta}$,
is positive definite by construction, it is also invertible for all times $t$. Then, multiplication of both sides of (\ref{eq:omega equation})
by $M\boldsymbol{\Theta}^{-1}$ gives the
closest rigid-body angular velocity

\begin{equation}
\boldsymbol{\omega}=M\boldsymbol{\Theta}^{-1}\overline{(\mathbf{x}-\bar{\mathbf{x}})\times(\mathbf{v}-\bar{\mathbf{v}})},\label{eq:optimal solution-1}
\end{equation}
a linear function of the velocity field $\mathbf{v}.$ 

In order to check whether the angular velocity vector defined by (\ref{eq:optimal solution-1})
is a minimizer of the functional $\widetilde{L_{1}}$, it is sufficient
to note that its Hessian is
\[
\frac{\partial^{2}}{\partial\omega_{n}\partial\omega_{p}}\widetilde{L_{1}}(\boldsymbol{\omega},t)=\frac{2}{M}[\boldsymbol{\Theta}]_{np},
\]
with the details of this calculation given in Appendix B.
By the positive definiteness of $\boldsymbol{\Theta}$ we
conclude that $\widetilde{L_{1}}(\boldsymbol{\omega},t)$ indeed has
a unique, strict global minimum at the rigid-body angular velocity 
$\boldsymbol{\omega}$ defined in (\ref{eq:optimal solution-1}). 

\subsection{Properties of the deformation velocity\label{subsec:Properties-of-the}}

\subsubsection{Basic properties of $\textbf{\ensuremath{\mathbf{v}_{d}}}$ }

We note that the deformation velocity is guaranteed to have no
further rigid-body velocity component under our decomposition principle.
Indeed, applying formulas (\ref{eq:optimal solution-1}) and (\ref{eq:definition of deformation velocity})
again to $\mathbf{v}=\textbf{\ensuremath{\mathbf{v}_{d}}}$ would
yield the secondary rigid-body angular velocity

\[
\boldsymbol{\omega}^{(2)}=\mathbf{0},
\]
as we show in detail in Appendix F. 

\subsubsection{Objectivity of $\textbf{\ensuremath{\mathbf{v}_{d}}}$\label{subsec:Objectivity-of-valpha} }

We recall that a vector field $\mathbf{a}(\mathbf{x},t)$ is said
to be \textit{objective} if under any Euclidean transformation
\begin{equation}
\mathbf{x}=\mathbf{Q}(t)\mathbf{y}+\mathbf{b}(t),\quad\mathbf{\mathbf{Q}\mathbf{Q}}^{T}=\mathbf{I},\quad\mathbf{Q}(t)\in\mathrm{SO}(3),\quad\mathbf{y},\mathbf{b}(t)\in\mathbb{R}^{3},\label{eq:Frame change}
\end{equation}
it transforms as 

\begin{equation}
\hat{\mathbf{a}}(\mathbf{y},t)=\mathbf{Q}^{T}(t)\mathbf{a}(\mathbf{x},t),\label{eq:objective vector field}
\end{equation}

In the transformed frame defined by (\ref{eq:Frame change}) the
transformed rigid-body angular velocity, $\boldsymbol{\omega}$,
satisfies 

\begin{equation}
\mathbf{\widehat{\boldsymbol{\Theta}}}\widehat{\boldsymbol{\omega}}=M\overline{(\mathbf{y}-\bar{\mathbf{y}})\times(\mathbf{\hat{v}}-\hat{\overline{\mathbf{v}}})},\label{eq:omegaOmega-1}
\end{equation}
with
\begin{equation}
\mathbf{\widehat{\boldsymbol{\Theta}}}=M\overline{\left(\left|\mathbf{y}-\bar{\mathbf{y}}\right|^{2}\right)\mathbf{I}-(\mathbf{y}-\bar{\mathbf{y}})\otimes(\mathbf{y}-\bar{\mathbf{y}})}.\label{eq:transformedI}
\end{equation}

The moment of inertia tensor $\boldsymbol{\Theta}$ 
is an objective tensor (\cite{Littlejohn97}), i.e., transforms
as
\begin{equation}
\mathbf{\widehat{\boldsymbol{\Theta}}}=\mathbf{Q}^{T}\boldsymbol{\Theta}\mathbf{Q}.\label{eq:transf of moment of inertia tensor}
\end{equation}
Differentiation of eq. (\ref{eq:Frame change}) with respect to time
gives the classic velocity transformation rule,
\begin{equation}
\begin{aligned}\widehat{\mathbf{v}} & =\mathbf{Q}^{T}\left(\mathbf{v}-\mathbf{\dot{\mathbf{Q}}}\mathbf{y}-\dot{\mathbf{b}}\right),\end{aligned}
\label{eq:transform v}
\end{equation}
which enables us to conclude in Appendix C  
 that the closest
rigid-body angular velocity transforms as 
\begin{equation}
\hat{\boldsymbol{\omega}}=\mathbf{Q}^{T}\left(\boldsymbol{\omega}-\dot{\mathbf{q}}\right),\label{eq:transform omega}
\end{equation}
where $\dot{\mathbf{q}}(t)$ is the dual of the skew-symmetric
matrix $\dot{\mathbf{\boldsymbol{Q}}}\mathbf{\boldsymbol{Q}}^{T}(t)$. 

In the $\mathbf{y}$-frame, the deformation velocity defined by eq.
(\ref{eq:definition of deformation velocity}) is simply
\begin{equation}
\hat{\mathbf{v}}_{d}=\mathbf{\hat{v}-\hat{\overline{v}}}-\mathbf{\boldsymbol{\widehat{\omega}}}\times(\mathbf{y}-\mathbf{\overline{y}}).\label{eq:tranformation of deformation velocity}
\end{equation}
Substituting the transformation rules (\ref{eq:transform v}) and
(\ref{eq:transform omega}) for $\mathbf{v}$ and $\boldsymbol{\omega}$
and $\mathbf{v}$ into eq. (\ref{eq:tranformation of deformation velocity}),
we then obtain in 
Appendix D that
\begin{equation}
\hat{\mathbf{v}}_{d}=\mathbf{Q}^{T}\mathbf{v_{\text{\ensuremath{d}}}}.\label{eq:deformation velocity is objective}
\end{equation}
 Therefore, by formula (\ref{eq:objective vector field}), the deformation
velocity $\mathbf{v}_{d}$ is objective. 

\subsubsection{Physical observability of $\textbf{\ensuremath{\mathbf{v}_{d}}}$ }

We now consider the specific frame change

\begin{equation}
\mathbf{x}=\mathbf{Q}_{RB}(t)\mathbf{y}+\mathbf{b}_{RB}(t),\label{eq:change to special observer}
\end{equation}
with $\mathbf{Q}_{RB}(t)$ and $\mathbf{b}_{RB}(t)$ defined
as solutions of the linear system of differential equations

\begin{align}
\dot{\mathbf{Q}}_{RB} & =\boldsymbol{\Omega}(t)\mathbf{Q}_{RB},\nonumber \\
\dot{\mathbf{b}}_{RB} & =\boldsymbol{\Omega}(t)\mathbf{b}_{RB}(t)+\mathbf{\bar{v}}(t)-\boldsymbol{\Omega}(t)\bar{\mathbf{x}},\label{eq:specific frame change}
\end{align}
satisfying the initial conditions $\mathbf{Q}_{RB}(t_{0})=\mathbf{I}$
and $\mathbf{b}_{RB}(t_{0})=\mathbf{0}$. In (\ref{eq:specific frame change}),
we used the tensorial form of the rigid body velocity, i.e., $\mathbf{\boldsymbol{\Omega}}$
is the dual of $\boldsymbol{\omega},$ as defined by
formula (\ref{eq:axvector}). Note that $\boldsymbol{\Omega}(t)$
is skew-symmetric and hence the fundamental solution of the first
equation in (\ref{eq:specific frame change}) is automatically proper
orthogonal. The second equation in (\ref{eq:specific frame change})
is also linear and hence has a unique solution $\mathbf{b}_{RB}(t)$
for the initial conditions specified. 

As we show in Appendix E, 
  under the observer change (\ref{eq:change to special observer}),
the full velocity field becomes

\begin{equation}
\hat{\mathbf{v}}=\mathbf{Q}_{RB}^{T}\textbf{\ensuremath{\mathbf{v}_{d}}}.\label{eq:v hat in special frame}
\end{equation}
At the same time, the general transformation formula (\ref{eq:deformation velocity is objective})
evaluated on the specific frame change (\ref{eq:change to special observer})
gives $\mathbf{\hat{v}}_{d}=\mathbf{Q}_{RB}^{T}\textbf{\ensuremath{\mathbf{v}_{d}}}$,
and hence eq. (\ref{eq:v hat in special frame}) implies

\begin{equation}
\hat{\mathbf{v}}=\mathbf{\hat{v}}_{d}=\mathbf{Q}_{RB}^{T}\textbf{\ensuremath{\mathbf{v}_{d}}}.\label{eq:velocity is the deformation velocity in hat frame}
\end{equation}

Therefore, the full velocity field coincides with its deformation
velocity component in the frame of the specific observer defined by
the linear ODEs in (\ref{eq:specific frame change}). This is because
the specific fame change (\ref{eq:specific frame change}) eliminates
both the mean velocity and the mean vorticity in the $\mathbf{y}$
frame, which makes the velocity field identical to its deformation
part. This special frame, therefore, is the frame
co-moving with the closest rigid-body velocity field $\mathbf{v}_{RB}$.

All this also implies that the trajectories of the ODE 
\begin{equation}
\dot{\mathbf{y}}=\mathbf{Q}_{RB}^{T}(t)\textbf{\ensuremath{\mathbf{v}_{d}}}(\mathbf{Q}_{RB}(t)\mathbf{y}+\mathbf{b}_{RB}(t),t)\label{eq:equivalent deformation velocity ODE}
\end{equation}
are mapped into the trajectories of the original velocity field $\mathbf{v}(\mathbf{\mathbf{x}},t)$
under the change of variables (\ref{eq:change to special observer}).
Therefore, the trajectories of (\ref{eq:equivalent deformation velocity ODE})
are smoothly conjugate of those of $\mathbf{v}(\mathbf{\mathbf{x}},t)$,
sharing the same dynamic features as trajectories of conjugate dynamical
systems always do (stability, asymptotic behavior, invariant manifolds
etc.). As any frame change, the transformation (\ref{eq:change to special observer})
also transforms originally fixed flow boundaries (as invariant manifolds
of the flow) into moving invariant surfaces. { Note, however, that there is, in general, no smooth conjugacy between the trajectories of $\mathbf{v}$ and those of $\mathbf{v}_d$. }

\subsection{Objectivization of physical fields using the deformation velocity}

The most broadly used Eulerian quantities (e.g., the kinetic energy,
momentum and vorticity) associated with a moving fluid depend on the
observer and hence do not capture purely intrinsic properties of the
flow. The objectivity of the deformation velocity $\textbf{\ensuremath{\mathbf{v}_{d}}}$,
however, makes it possible to eliminate this observer dependence by
simply computing these classic Eulerian quantities from $\textbf{\ensuremath{\mathbf{v}_{d}}}$
as opposed to $\mathbf{v}$. In view of formula (\ref{eq:velocity is the deformation velocity in hat frame}),
the replacement of $\mathbf{v}$ with $\textbf{\ensuremath{\mathbf{v}_{d}}}$
is equivalent to evaluating the original non-objective quantities
in a special, objectively determined frame that is uniquely identifiable
by any physical observer.

Specifically, the pointwise \emph{deformation kinetic energy}, $E_{d}(\mathbf{x},t)$,
\emph{deformation enstrophy}, $\mathcal{E}_{d}(\mathbf{x},t)$,
and \emph{deformation helicity}, $H_{d}(\mathbf{x},t)$, can
be defined by evaluating their classic counterparts, $E=\frac{1}{2}\left|\mathbf{v}\right|^{2}$,
$\mathcal{E}=\left|\boldsymbol{\nabla}\times\textbf{\ensuremath{\mathbf{v}}}\right|^{2}$
and $H=\left|\textbf{\ensuremath{\mathbf{v}}}\cdot\boldsymbol{\nabla}\times\textbf{\ensuremath{\mathbf{v}}}\right|^{2}$,
in the frame co-moving with $\mathbf{v}_{RB}(\mathbf{x},t)$, as defined
in eq. (\ref{eq:change to special observer}). The scalar fields
\begin{align}
E_{d}(\mathbf{x},t) & =\frac{1}{2}\left|\hat{\mathbf{v}}(\mathbf{y},t)\right|^{2}=\frac{1}{2}\left|\textbf{\ensuremath{\mathbf{v}_{d}}}(\mathbf{x},t)\right|^{2},\nonumber \\
\mathcal{E}_{d}(\mathbf{x},t) & =\left|\hat{\boldsymbol{\nabla}}\times\hat{\mathbf{v}}(\mathbf{y},t)\right|^{2}=\left|\boldsymbol{\nabla}\times\textbf{\ensuremath{\mathbf{v}_{d}}}(\mathbf{x},t)\right|^{2},\nonumber \\
H_{d}(\mathbf{x},t) & =\left|\hat{\mathbf{v}}(\mathbf{y},t)\cdot\hat{\boldsymbol{\nabla}}\times\hat{\mathbf{v}}(\mathbf{y},t)\right|^{2}=\left|\textbf{\ensuremath{\mathbf{v}_{d}}}(\mathbf{x},t)\cdot\boldsymbol{\nabla}\times\textbf{\ensuremath{\mathbf{v}_{d}}}(\mathbf{x},t)\right|^{2},\label{eq:objectivized scalars}
\end{align}
are objective, returning pointwise energy, enstrophy and helicity
values that are independent of the observer. This also means that
the various topological features (such as the level sets) of these
fields are indifferent to changes in the observer and hence are intrinsic
to the fluid, unlike those of $E(\mathbf{x},t)$, $\mathcal{E}(\mathbf{x},t)$,
and $H(\mathbf{x},t)$. The spatial integrals of the Eulerian scalar
fields in (\ref{eq:objectivized scalars}) over the flow domain $U$
are also frame-invariant, given the pointwise objectivity of their
integrands. 

As an exception, the turbulent kinetic energy, 
\[
k(\mathbf{\mathbf{x}},t)=\frac{1}{T}\int_{t}^{t+T}\left|\mathbf{v}(\mathbf{\mathbf{x}},\tau)-\frac{1}{T}\int_{\tau}^{\tau+T}\mathbf{v}(\mathbf{\mathbf{x}},s)\,ds\right|^{2}d\tau,
\]
whose definition relies on the assumption of a well-defined temporal
mean at each point of the flow (\cite{pope_turbulent_2000}), does not
become objective even when evaluated on $\textbf{\ensuremath{\mathbf{v}_{d}}}$.
Indeed, in a transformed frame, we would have 
\begin{align*}
\hat{k}_{d} & =\frac{1}{T}\int_{t}^{t+T}\left|\hat{\mathbf{v}}_{d}(\mathbf{y},\tau)-\frac{1}{T}\int_{\tau}^{\tau+T}\hat{\mathbf{v}}_{d}(\mathbf{y},s)ds\right|^{2}d\tau\\
 & =\frac{1}{T}\int_{t}^{t+T}\left|\mathbf{Q}^{T}(\tau)\textbf{\ensuremath{\mathbf{v}_{d}}}(\mathbf{y},\tau)-\frac{1}{T}\int_{\tau}^{\tau+T}\mathbf{Q}^{T}(s)\textbf{\ensuremath{\mathbf{v}_{d}}}(\mathbf{y},s)ds\right|^{2}d\tau\\
 & =\frac{1}{T}\int_{t}^{t+T}\left|\textbf{\ensuremath{\mathbf{v}_{d}}}(\mathbf{y},\tau)-\frac{1}{T}\mathbf{Q}(\tau)\int_{\tau}^{\tau+T}\mathbf{Q}^{T}(s)\textbf{\ensuremath{\mathbf{v}_{d}}}(\mathbf{y},s)ds\right|^{2}d\tau\\
 & \neq k_{d}.
\end{align*}

In contrast, the \emph{deformation momentum $\mathbf{p}_{d}$
}and the\emph{ deformation vorticity} $\mathbf{w}_{d}$, defined
as
\begin{align*}
\mathbf{p}_{d}(\mathbf{x},t) & =\rho(\mathbf{x},t)\mathbf{v}_{d}(\mathbf{x},t),\\
\mathbf{w}_{d}(\mathbf{x},t) & =\boldsymbol{\nabla}\times\textbf{\ensuremath{\mathbf{v}_{d}}}(\mathbf{x},t),
\end{align*}
become objective vector fields. This follows from the objectivity
of $\textbf{\ensuremath{\mathbf{v}_{d}}}$ established in formula
(\ref{eq:deformation velocity is objective}). Likewise, the \emph{deformation
velocity gradient}, $\boldsymbol{\nabla}\textbf{\ensuremath{\mathbf{v}_{d}}}$,
the \emph{deformation rate-of-strain tensor} $\textbf{\ensuremath{\mathbf{S}_{d}}}=\frac{1}{2}\left(\boldsymbol{\nabla}\mathbf{v}_{d}+\left[\boldsymbol{\nabla}\mathbf{v}_{d}\right]^{T}\right)$
and the \emph{deformation spin tensor,} ${\textbf{\ensuremath{\mathbf{W}_{d}}}=\frac{1}{2}\left(\boldsymbol{\nabla}\mathbf{v}_{d}-\left[\boldsymbol{\nabla}\mathbf{v}_{d}\right]^{T}\right)}$
are all objective tensors. In addition, we have 
\begin{equation}
\textbf{\ensuremath{\mathbf{S}_{d}}}(\mathbf{\mathbf{x}},t)\equiv\textbf{\ensuremath{\mathbf{S}}}(\mathbf{\mathbf{x}},t),\label{eq:rate of strain preserved}
\end{equation}
 where $\textbf{\ensuremath{\mathbf{S}}}=\frac{1}{2}\left(\boldsymbol{\nabla}\mathbf{v}+\left[\boldsymbol{\nabla}\mathbf{v}\right]^{T}\right)$
is the classic rate-of-strain tensor. 

A consequence of eq. (\ref{eq:rate of strain preserved}) is that
the deformation velocity also contains all objective Eulerian coherent
structures (OECS; see \cite{serra_objective_2016}) of the original velocity
field. These coherent structures are defined as the instantaneous
limits of Lagrangian coherent structures (LCSs; see \cite{haller_lagrangian_2015})
and hence govern the advection of material fluid elements for short
times. Passage to the deformation velocity also preserves another
Eulerian indicator of coherence, the instantaneous vorticity deviation
${\text{IVD}(\mathbf{x},t)=\frac{1}{2}\left|\mathbf{w}(\mathbf{x},t)-\overline{\mathbf{\mathbf{w}}(\mathbf{x},t)}\right|}$,
defined by \cite{haller_defining_2016}. Indeed, 
\begin{align*}
\text{IVD}_{d} & =\frac{1}{2}\left|\mathbf{w}_{d}-\overline{\mathbf{\mathbf{w}}_{d}}\right|=\frac{1}{2}\left|\mathbf{w}-2\boldsymbol{\omega}-\overline{\mathbf{\mathbf{w}}_{d}}+2\boldsymbol{\omega}\right|=\text{IVD}.
\end{align*}

More generally, even though they often lack a rigorous connection
to the flow, all the classic vortex criteria reviewed in \cite{kolar_vortex_2007},  \cite{epps_review_2017}, \cite{gunther_state_2018}, and  \cite{haller_can_2021}
become indifferent to the observer when evaluated on $\mathbf{v}_{d}$,
or, equivalently, on the full velocity field in a frame co-moving
with $\mathbf{v}_{RB}$. 

{ As a Lagrangian example, we recall the trajectory length function proposed by \cite{Mancho2013}, defined as \begin{equation}\label{eq:mfunc}
    M_{t_0}^{t_1}(\mathbf{x}_0) = \int_{t_0}^{t_1} |\mathbf{v}(\mathbf{x}(s; \mathbf{x}_0),s)|ds. 
\end{equation} 
Evaluating \eqref{eq:mfunc} on $\mathbf{v}_d$ makes $M_{t_0}^{t_1}$ objective. Indeed, as $\hat{\mathbf{v}}_d = \mathbf{Q}^T \mathbf{v}_d$, the transformed trajectory length is 
\begin{equation}\label{eq:mfunc}
    \hat{M}_{t_0}^{t_1}(\mathbf{y}_0) = \int_{t_0}^{t_1} |\hat{\mathbf{v}}_d(\mathbf{y}(s; \mathbf{y}_0),s)|ds = M_{t_0}^{t_1}(\mathbf{x}_0). 
\end{equation} }
\section{Examples\label{sec:Examples}}

We now briefly illustrate the extraction of deformation velocities
in two-dimensional and three-dimensional examples.

\subsection{Deformation velocities in 2D}

First, we perform the proposed velocity decomposition on two of the
explicit two-dimensional unsteady Navier--Stokes solutions derived
in \cite{pedergnana_explicit_2020}. Our first example is the velocity
field
\begin{equation}
\mathbf{v}(\mathbf{x},t)=\left(\begin{array}{c}
\dot{x}\\
\dot{y}
\end{array}\right)=\left(\begin{array}{cc}
\sin4t & \cos4t+2\\
\cos4t-2 & -\sin4t
\end{array}\right)\left(\begin{array}{c}
x\\
y
\end{array}\right)+0.005\left(\begin{array}{c}
x(x^{2}-3y^{2})\\
-x(3x^{2}-y^{2})
\end{array}\right).\label{eq:2dsystem}
\end{equation}

For an arbitrary length parameter $a>0$, we consider square-shaped
flow domains of the form $U_{a}=\left\{ (x,y)\in\mathbb{R}^{2}:\,\,\,\,\,(x,y)\in[-a,a]^{2}\right\} $,
on which to find the deformation velocity component $\mathbf{v}_{d}(\mathbf{x},t)$
of (\ref{eq:2dsystem}). As our optimization problem (\ref{eq:functionalSpecific})
was posed for three-dimensional velocity fields, we formally extend
(\ref{eq:2dsystem}) with an identically zero velocity component in
the $z$-direction. As a result, the angular velocity vector $\boldsymbol{\omega}$,
defined by (\ref{eq:optimal solution-1}), will only have a single
nonzero component, $\omega_{d}=\left(\boldsymbol{\omega}\right)_{z}$.
In this simple example, an analytic computation of $\omega_{d}$
is possible, yielding $\omega_{d}=-2$. 

In fig. \ref{fig:fig1}a, we show the streamlines of system (\ref{eq:2dsystem})
for the time instant $t=0$ (other time instants yield similar streamlines)
over the domain $U_{1}$. In fig. \ref{fig:fig1}b, we show the instantaneous
streamlines of the deformation velocity field $\mathbf{v}_{d}$
obtained from formulas (\ref{eq:optimal solution})-(\ref{eq:definition of deformation velocity})
over $U_{1}$. In fig. \ref{fig:fig1}c-d, we again show the original
and deformation streamlines for the larger domain $U_{30}$.

The streamline structure of the original velocity field (\ref{eq:2dsystem})
suggests a vortical region around the origin, which is also the prediction
of all classic, frame-dependent vortex criteria when applied to this
example. In contrast, \cite{pedergnana_explicit_2020} show that the
origin is a saddle-type LCS residing in a hyperbolic region that exhibits
chaotic mixing. This behavior makes (\ref{eq:2dsystem}) a false positive
of the traditional vortex criteria. In spite of this, instantaneous
streamlines of the deformation velocity correctly reveal the saddle
type of the origin, as seen in fig. \ref{fig:fig1}b. We recall that
the instantaneous streamlines have generally no direct connection
with the Lagrangian dynamics in an unsteady flow and this fact remains
true for the deformation velocity $\mathbf{v}_{d}$ as well.
Yet, in this example, the objective streamlines of $\mathbf{v}_{d}$
correctly reflect the Lagrangian stability properties of the origin. 

\begin{figure}
\includegraphics[width=1\textwidth]{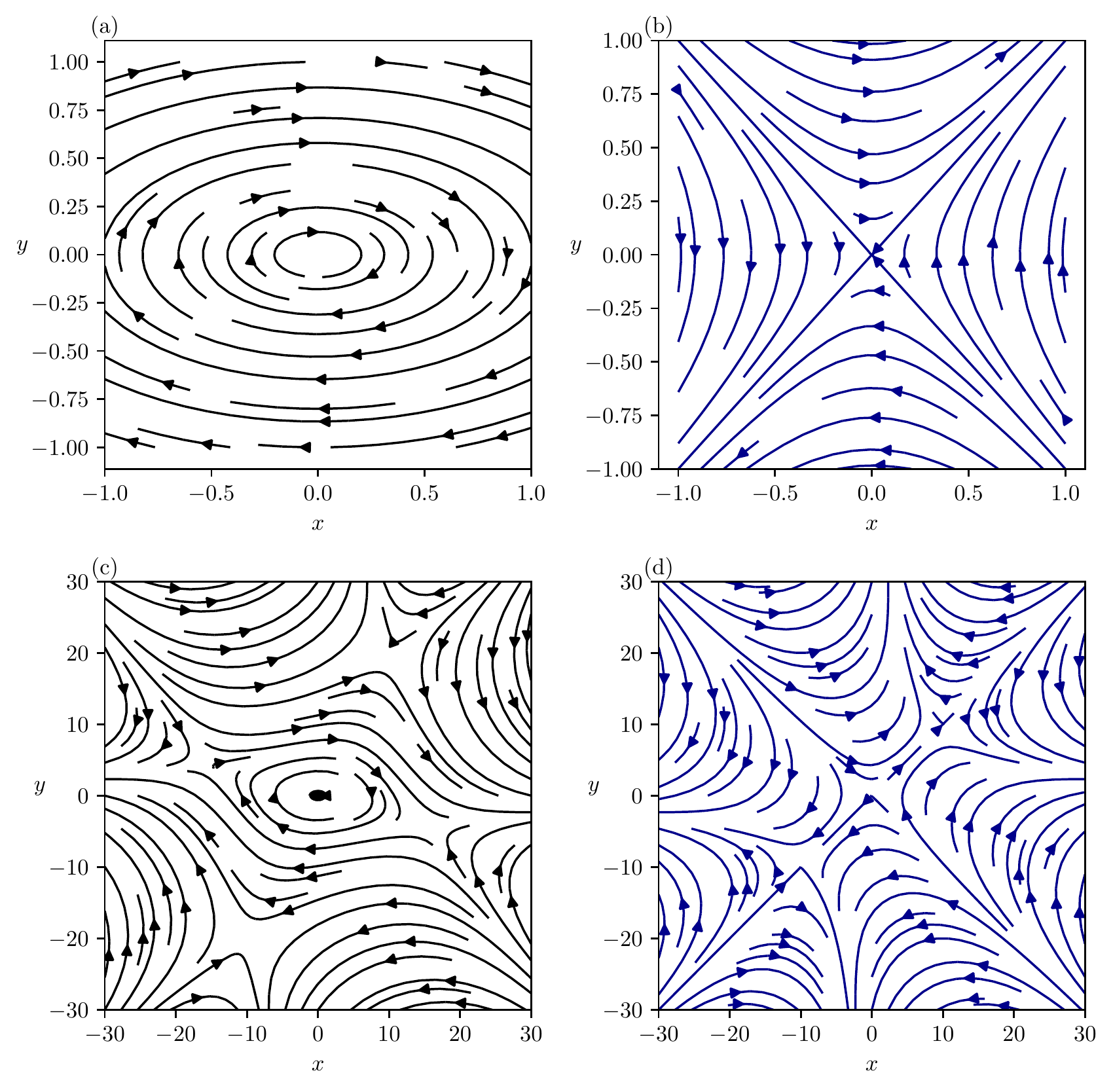}\caption{\label{fig:fig1}Left column: Observer-dependent streamlines of the
velocity field (\ref{eq:2dsystem}) over the domains $U_{1}$ (a)
and $U_{30}$ (c). Right column: Objective streamlines of the corresponding
deformation velocity field $\mathbf{v}_{d}(\mathbf{x},t)$ over
the domains $U_{1}$ (b) and $U_{30}$ (d). }
\end{figure}

Our second example, taken from the same class of explicit Navier--Stokes
solutions as (\ref{eq:2dsystem}), is the velocity field

\begin{equation}
\mathbf{v}(\mathbf{x},t)=\left(\begin{array}{c}
\dot{x}\\
\dot{y}
\end{array}\right)=\left(\begin{array}{cc}
\sin4t & \cos4t+\frac{1}{2}\\
\cos4t-\frac{1}{2} & -\sin4t
\end{array}\right)\left(\begin{array}{c}
x\\
y
\end{array}\right)-0.015\left(\begin{array}{c}
x^{2}-y^{2}\\
-2xy
\end{array}\right).\label{eq:2dsystemB}
\end{equation}

In \cite{pedergnana_explicit_2020}, (\ref{eq:2dsystemB}) is shown
to be a false negative for all available frame-dependent vortex criteria.
Indeed, the instantaneous streamlines indicate a saddle-type structure
near the origin, yet the Lagrangian particle motion is quasiperiodic
(elliptic). In fig. \ref{fig:fig2}, we show the streamlines of the
velocity field (\ref{eq:2dsystemB}), along with the streamlines of
its deformation component, which has $\omega=-\frac{1}{2}$. This time, the streamlines of the deformation
component have the same structure as those of the original velocity
field. In panel (c) of fig. \ref{fig:fig2}, we show the velocity
field, as observed from the frame co-rotating with the closest rigid
body velocity. This is calculated by solving for $\mathbf{Q}_{RB}(t)$
according to (\ref{eq:specific frame change}). In this case, the
solution can be written out explicitly,
\[
\mathbf{Q}_{RB}(t)=\left(\begin{array}{cc}
\cos\omega t & -\sin\omega t\\
\sin\omega t & \cos\omega t
\end{array}\right),
\]
as noted for two-dimensional flows in \cite{haller_can_2021}. The
velocity field observed in the frame co-rotating with $\mathbf{v}_{RB}(\mathbf{x},t)$
is (\ref{eq:equivalent deformation velocity ODE}). The trajectories
of $\mathbf{Q}_{RB}^{T}(t)\mathbf{v}_{d}(\mathbf{x},t)$
are indeed topologically equivalent to those of $\mathbf{v}.$ This
is illustrated in panel (d) of fig. \ref{fig:fig2}, where a trajectory
of $\mathbf{v}$ is shown together with the corresponding trajectory
of $\mathbf{Q}_{RB}^{T}(t)\mathbf{v}_{d}(\mathbf{x},t)$. 

\begin{figure}

\includegraphics[width=1\textwidth]{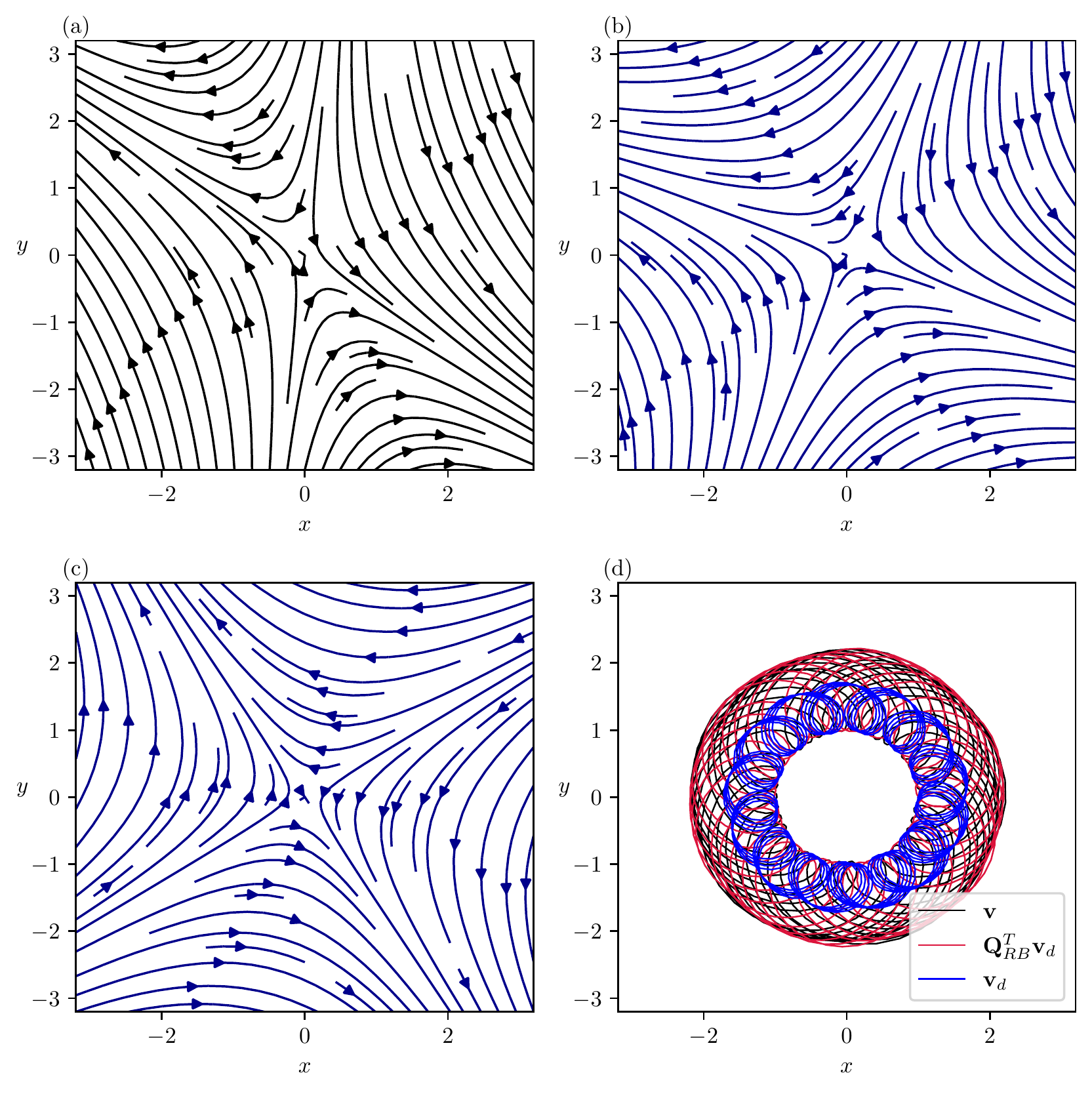}\caption{\label{fig:fig2} The velocity field (\ref{eq:2dsystemB}) and its
deformation velocity component. (a) Streamlines of (\ref{eq:2dsystemB})
at time $t=10$. Panel (b) shows the streamlines of the deformation
component $\mathbf{v}_{d}$, while panel (c) shows the streamlines
of the observed velocity, $\mathbf{Q}_{RB}^{T}(t)\mathbf{v_{d}}$.
In panel (d), a trajectory of the velocity field $\mathbf{v}$ is
shown in black. The corresponding trajectory of the deformation
component, $\mathbf{v}_{d}$, (of the observed velocity, $\mathbf{Q}_{RB}^{T}(t)\mathbf{v}_{d}$)
is shown in blue (red). }
\end{figure}

Our third example is a kinematic velocity model describing an unsteady
gyre in a rotating circular tank, with free slip boundary conditions
along the tank wall (\cite{lekien_unsteady_2008}). This unsteady flow
develops Lagrangian flow separation and reattachment, exhibited by
a sharp material spike emanating from the boundary and hitting the
boundary again at a diametrically opposite point. This separation
and reattachment, however, have no indication in the Eulerian frame
as all streamlines remain circular for all times (see fig. \ref{fig:fig3}a).
Although \cite{lekien_unsteady_2008} show that the Lagrangian separation
does not occur at instantaneous stagnation points on the boundary,
such points often serve as rough indicators of flow separation (\cite{zhu_numerical_2021}).
Their absence, therefore, suggests a lack of flow separation even
though that is not the case here.

The two-dimensional model velocity field in \cite{lekien_unsteady_2008}
has the stream function
\begin{equation}
\Psi(\mathbf{x},t)=(|\mathbf{x}|^{2}-1)(x\sin\omega_{s}t+y\cos\omega_{s}t)-\frac{1}{2}\omega_{s}|\mathbf{x}|^{2}.\label{eq:streamfn}
\end{equation}
\sloppy Our physical domain is the rotating circular container, i.e., ${U=\left\{ (x,y)\in\mathbb{R}^{2}:\,\,\,\,\,x^{2}+y^{2}\leq1\right\} }$, on which we recover
the rigid body angular velocity $\omega=\omega_{s}$, as anticipated. In fig. \ref{fig:fig3}b,
we show the instantaneous streamlines of the deformation velocity
$\mathbf{v}_{d}$ at the same time instant that we used for plotting
the streamlines of $\mathbf{v}$ in fig. \ref{fig:fig3}a. We note
the appearance of two, diametrically opposite stagnation points for
$\mathbf{v}_{d}$ along the boundary, correctly suggesting the
simultaneous presence of separation and reattachment in the flow.

\begin{figure}
\centering{}\includegraphics[width=1\textwidth]{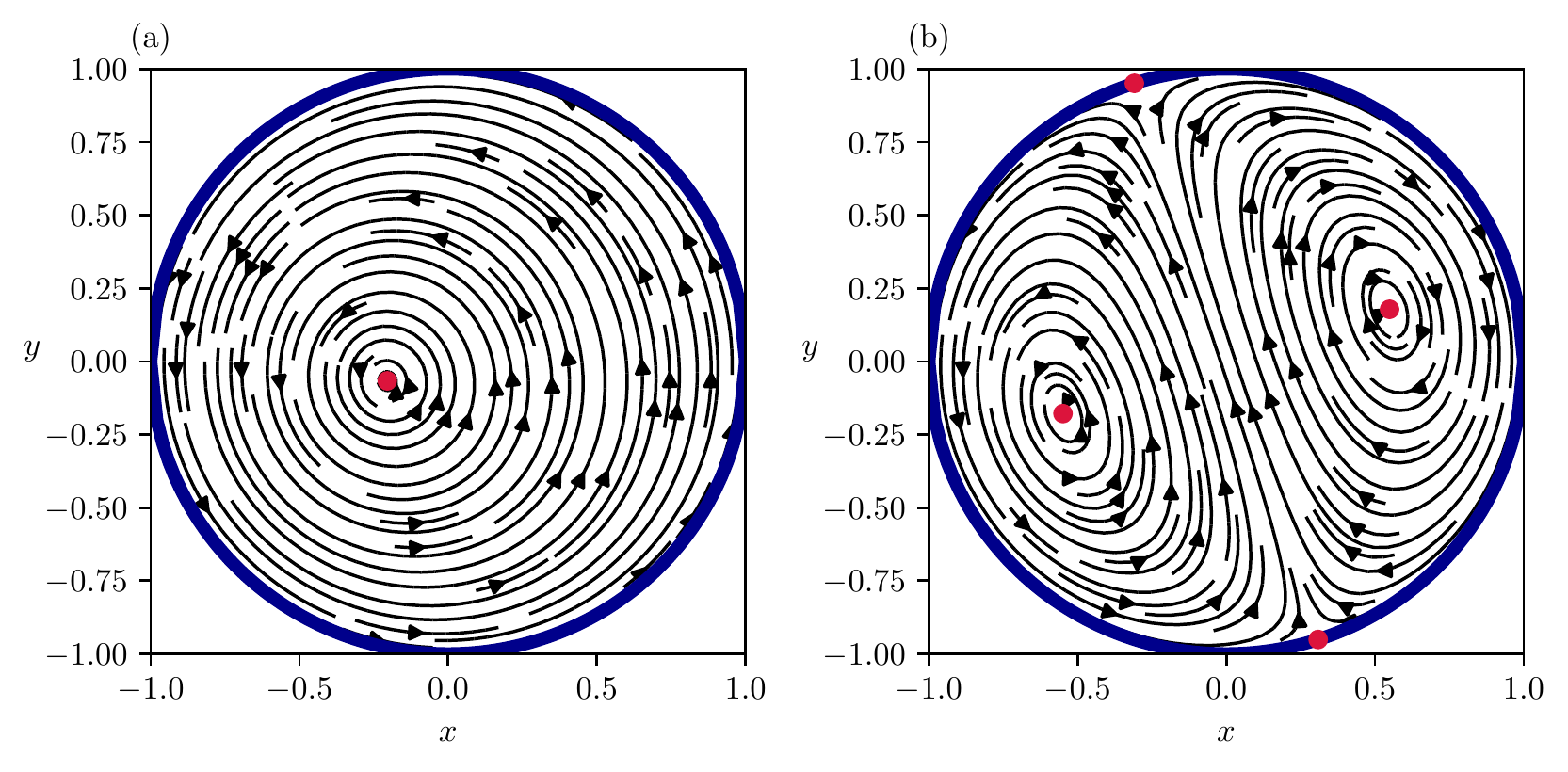}\caption{\label{fig:fig3}Streamlines of the kinematic model (\ref{eq:streamfn})
(a) and those of its deformation velocity component (b). The boundary
of the domain $U$ is shown in blue, while red points denote the stagnation
points in the flow. For both panels, $t=\pi/10$. The rigid body angular
velocity is $\omega=\omega_{s}=4$. }
\end{figure}

As a fourth example, we illustrate the calculation of the deformation
velocity filed on a two-dimensional unsteady velocity data set from
AVISO satellite altimetry measurements (\cite{le_traon_improved_1998}).
We work with a single velocity snapshot from April 12, 2014. The
domain of the dataset is the Gulf Stream region. The velocity field
is obtained from the sea surface height (SSH), which is the stream
function of the surface velocity field under the geostrophic assumption.

{ By choosing the domain to be too large, we might include regions in which multiple eddies coexist. To highlight this dependence of $\mathbf{v}_d$ on the domain of interest, we perform the $\mathbf{v}=\mathbf{v}_{RB}+\mathbf{v}_{d}$ decomposition
of the surface velocity field over two different domains. First, we select the full domain $U=[290^{o},310^{o}]\times[33^{o},41^{o}]$
in longitude-latitude coordinates. In this case, we obtain the small, negative value $\omega=-0.0037$.}
In contrast, if we choose $U$ around an Eulerian mesoscale eddy highlighted
in fig. \ref{fig:fig4}b, then $\omega=0.6995$ is a considerably
larger positive number, indicative of the closest rigid body rotation
well describing the eddy.
This was to be expected since the full observed
region appears to contain several rotating features and hence lacks
a single dominant rigid-body rotation component. Indeed, a comparison
of panels (a), (b) confirms that the subtraction of the rigid
body velocity field does not alter the structure of the streamlines
considerably on this large domain. {As a result, the objectivized scalar fields such as the kinetic energy would also be practically indistinguishable from those evaluated on the original velocity field. 

For practical purposes, this means that when working with velocity fields coming from models or observations that cover a large domain of the globe, one has to first isolate smaller regions of interest. Otherwise, for large domains, the principle outlined here returns that there is no single co-moving observer and we have $\mathbf{v}_{RB}\approx 0$.
}

In contrast, the velocity field in the region around the eddy of panel
(a) is expected to have a dominant rigid-body component. This is consistent with the observations of \cite{tel2018}, who showed experimentally that the core of a vortex approximately rotates as a rigid body. Nevertheless, we find that the
deformation component is still nonzero. The inset
of panel (b) shows the streamlines of the deformation
component computed with respect to this eddy. 

\begin{figure}
\centering
\includegraphics[width=0.8\textwidth]{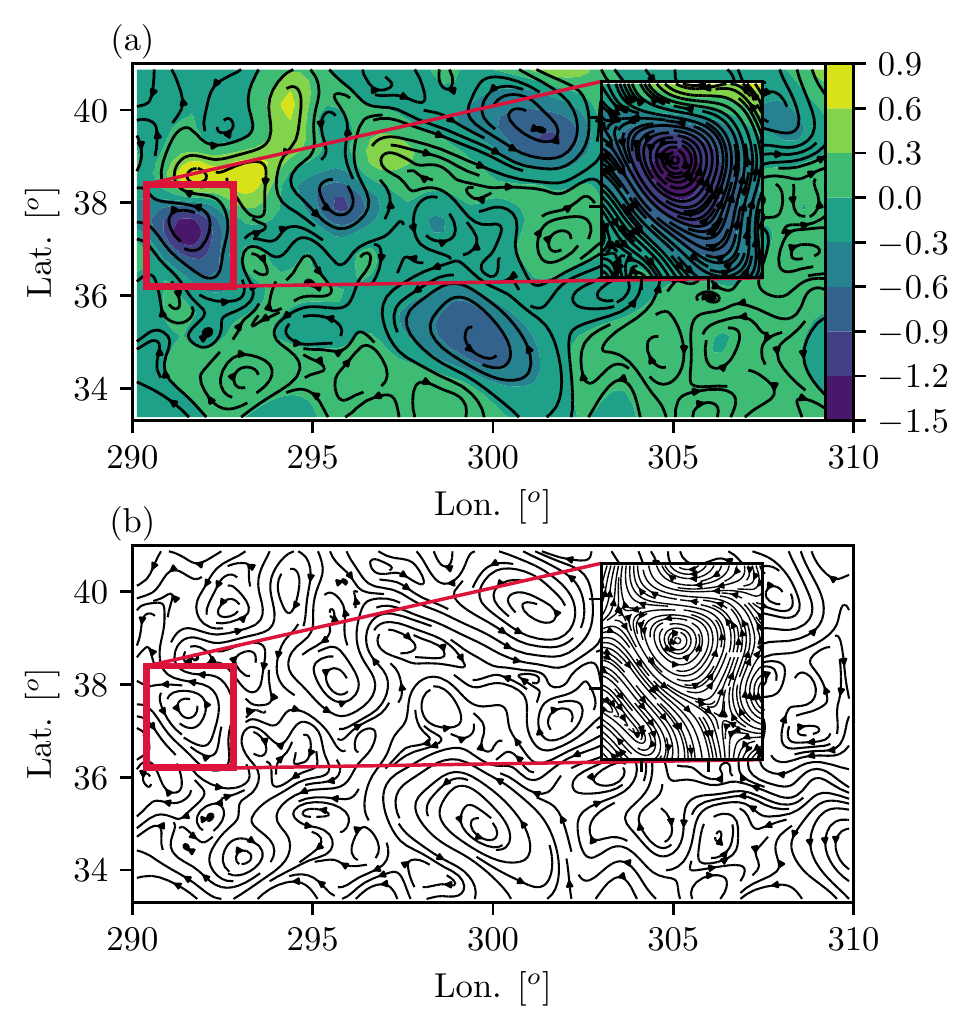}

\caption{\label{fig:fig4} The AVISO velocity field and its objective deformation
component on April 12, 2014. (a) The original streamlines overlaid
on the sea surface height (SSH) field measured in meters. The inset
shows a smaller subdomain encircling a mesoscale eddy. (b): objective
streamlines of the deformation velocity $\mathbf{v}_{d}$. The inset in (b) shows the streamlines of the deformation component calculated from the small region around the eddy.}
\end{figure}

\subsection{Objectivized Eulerian fields in 3D}

We now discuss the following objectivized Eulerian scalar fields:
the deformation kinetic energy $E_{d}(\mathbf{x}),$ the deformation
enstrophy $\mathcal{E}_{d}(\mathbf{x})$, and the deformation
helicity $H_{d}(\mathbf{x})$ defined in (\ref{eq:objectivized scalars}).
We calculate them for the classic ABC velocity field of \cite{dombre_chaotic_1986},
given by

\begin{equation}
\mathbf{v}(\mathbf{x},t)=\begin{pmatrix}\dot{x_{1}}\\
\dot{x_{2}}\\
\dot{x_{3}}
\end{pmatrix}=\left(\begin{array}{c}
A\sin x_{3}+C\cos x_{2}\\
B\sin x_{1}+A\cos x_{3}\\
C\sin x_{2}+B\cos x_{1}
\end{array}\right).\label{eq:abc}
\end{equation}

We take the domain $U=[-\pi,\pi]^{3}$ for the calculation of the
deformation component. After applying formula (\ref{eq:optimal solution-1}),
we obtain 
\begin{equation}
\bar{\mathbf{x}}=\mathbf{0},\qquad\bar{\mathbf{v}}=\mathbf{0},\qquad\boldsymbol{\omega}=\frac{3}{2\pi^{2}}\begin{pmatrix}C\\
A\\
B
\end{pmatrix},\label{eq:abcomega}
\end{equation}
yielding the deformation velocity
\[
\mathbf{v}_{d}=\left(\begin{array}{c}
A\sin x_{3}+C\cos x_{2}\\
B\sin x_{1}+A\cos x_{3}\\
C\sin x_{2}+B\cos x_{1}
\end{array}\right)-\frac{3}{2\pi^{2}}\begin{pmatrix}Bx_{2}-Ax_{3}\\
Cx_{3}-Bx_{1}\\
Ax_{1}-Cx_{2}
\end{pmatrix}.
\]

In fig. \ref{fig:fig5}, we show level sets of the energy, half the
enstrophy, and half the helicity of the ABC flow with the classic
choice of parameters $A=\sqrt{3}$, $B=\sqrt{2}$, $C=1$, which results
in chaotic Lagrangian dynamics (\cite{dombre_chaotic_1986}). For comparisons,
we also show the objectivized versions of the same quantities, calculated
from the deformation velocity (\ref{eq:definition of deformation velocity}),
with $\omega$ given by (\ref{eq:abcomega}). Since for the
ABC flow, $\mathbf{v}\equiv\nabla\times\mathbf{v},$ the classic enstrophy
and the helicity are equal to twice the energy. This, however, is
no longer the case for the deformation velocity, for which we have
\[
\nabla\times\mathbf{v}_{d}=\nabla\times\mathbf{v}-\nabla\times\left(\boldsymbol{\omega}\times\mathbf{x}\right)=\mathbf{v}-2\boldsymbol{\omega}.
\]
As a consequence, the deformation kinetic energy is not proportional
to either the deformation enstrophy or the deformation helicity. In
addition, the latter three quantities are objective scalar fields
while their counterparts, $E$, $\mathcal{E}$, and $H$ depend on
the observer.

\begin{figure}
\centering{}\includegraphics[width=1\textwidth]{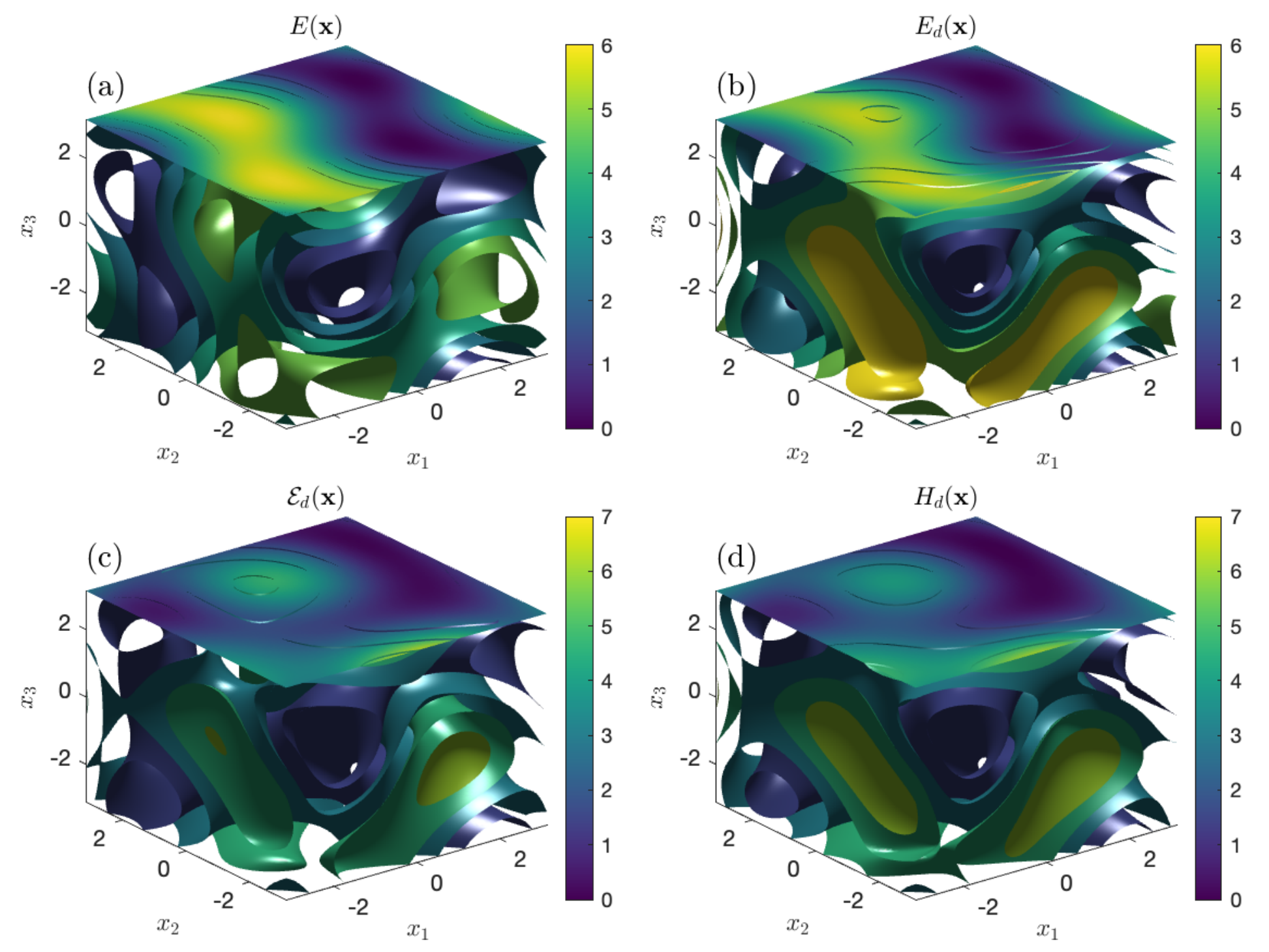}\caption{\label{fig:fig5}Energy, enstrophy and helicity isosurfaces for the
ABC flow. In panel (a), isosurfaces of the energy $E=\frac{1}{2}|\mathbf{v}(\mathbf{x},t)|^{2}$
are shown for the velocity field (\ref{eq:abc}), with parameters
$A=\sqrt{3}$, $B=\sqrt{2}$, $C=1$. These coincide with isosurfaces
of both $\frac{1}{2}\mathcal{E}=\frac{1}{2}\left|\boldsymbol{\nabla}\times\textbf{\ensuremath{\mathbf{v}}}\right|^{2}$
and $\frac{1}{2}H=\frac{1}{2}\left|\mathbf{v}\cdot\boldsymbol{\nabla}\times\textbf{\ensuremath{\mathbf{v}}}\right|^{2}$.
Panel (b) shows isosurfaces of the deformation kinetic energy $E_{d}=\frac{1}{2}|\mathbf{v}_{d}(\mathbf{x},t)|^{2}$.
In panels (c) and (d), isosurfaces of the deformation enstrophy, $\frac{1}{2}\mathcal{E}_{d}=\frac{1}{2}\left|\boldsymbol{\nabla}\times\textbf{\ensuremath{\mathbf{v}}}_{d}\right|^{2}$
and the deformation helicity, $\frac{1}{2}H_{d}=\frac{1}{2}\left|\mathbf{v}_{d}\cdot\boldsymbol{\nabla}\times\textbf{\ensuremath{\mathbf{v}}}_{d}\right|^{2}$
are shown, respectively.
The deformation velocity field $\mathbf{v}_{d}(\mathbf{x},t)$
is given as $\mathbf{v}_{d}(\mathbf{x},t)=\mathbf{v}(\mathbf{x},t)-\boldsymbol{\omega}\times\mathbf{x}$,
where $\mathbf{v}$ and $\text{\ensuremath{\boldsymbol{\omega}}}$
are defined in (\ref{eq:abc}) and (\ref{eq:abcomega}).}
\end{figure}

To illustrate the objectivity of the deformation kinetic energy $E_{d}$,
we transform the velocity field (\ref{eq:abc}) to a frame rotating
around the $z-$ axis with a constant angular velocity $\omega_{r}=5$, defined by $\mathbf{x=\mathbf{Q}}(t)\mathbf{y}$,
where 
\begin{equation}
\mathbf{Q}(t)=\left(\begin{array}{ccc}
\cos\omega_{r}t & -\sin\omega_{r}t & 0\\
\sin\omega_{r}t & \cos\omega_{r}t & 0\\
0 & 0 & 1
\end{array}\right).\label{eq:rotatingFrameABC}
\end{equation}

The velocity of the ABC flow observed in the $\mathbf{y}$-frame is
then
\begin{equation}
\dot{\mathbf{y}}=\mathbf{\hat{v}}(\mathbf{y},t)=\mathbf{Q}^{T}(t)\left(\mathbf{v}(\mathbf{Q}(t)\mathbf{y})-\dot{\mathbf{Q}}(t)\mathbf{y}\right).\label{eq:transformed velocity}
\end{equation}

After substituting (\ref{eq:abc}) and the expression for $\mathbf{Q}(t)$
into the transformation formula (\ref{eq:transformed velocity}),
we obtain an explicitly time dependent velocity field $\mathbf{\hat{v}}(\mathbf{y},t)$.
As discussed in Section \ref{subsec:Objectivity-of-valpha}, we then apply
the formulas (\ref{eq:optimal solution}) and (\ref{eq:definition of deformation velocity})
to calculate the deformation velocity $\widehat{\mathbf{v}}_{d}(\mathbf{y},t)$
in the $\mathbf{y}$-frame. In fig. \ref{fig:fig6}, we show the kinetic
energies $E$ and $\hat{E}$ calculated in the $\mathbf{x}$- and
$\mathbf{y}$- frames, respectively, along with their deformation
counterparts, $E_{d}$ and $\hat{E}_{d}$.

\begin{figure}
\includegraphics[width=\textwidth]{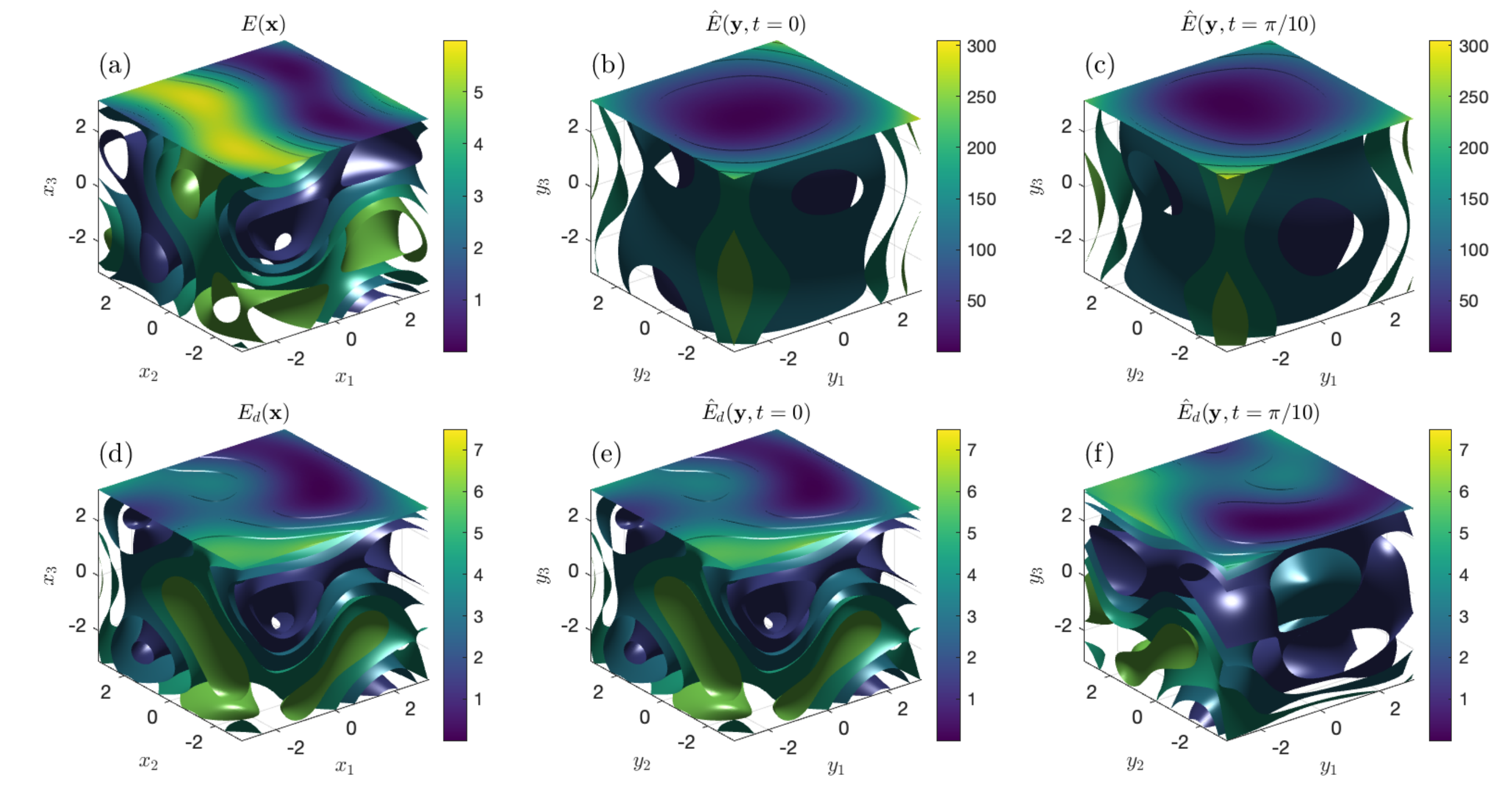}\caption{\label{fig:fig6}The kinetic energy $E$ and deformation kinetic energy
$E_{d}$ of the ABC flow observed in different frames. In panel
(a), isosurfaces of the energy $E=\frac{1}{2}|\mathbf{v}(\mathbf{x},t)|^{2}$
are shown for the velocity field (\ref{eq:abc}), with parameters
$A=\sqrt{3}$, $B=\sqrt{2}$, $C=1$. In panels (b) and (c) the isosurfaces
of the transformed kinetic energy, $\widehat{E}=\frac{1}{2}|\mathbf{\hat{v}}(\mathbf{y},t)|^{2}$,
are shown for $t=0$ and $t=\pi/10$. The transformation between the
$\mathbf{x}$ and $\mathbf{y}$ frames is $\mathbf{x}=\mathbf{Q}(t)\mathbf{y}$,
with $\mathbf{Q}(t)$ given by (\ref{eq:rotatingFrameABC}). Coordinates in the $\mathbf{y}-$ frame are labeled as ($y_1$, $y_2$, $y_3$). In panel
(d) the isosurfaces of the deformation kinetic energy $E_{d}(\mathbf{x})$
are shown, while panels (e)-(f) show the isosurfaces of
$\hat{E}_{d}(\mathbf{y},t)$, the deformation kinetic energy
computed in the $\mathbf{y}$-frame for $t=0$ and $t=\pi/10$. }

\end{figure}

When observed from the frame defined by (\ref{eq:rotatingFrameABC}),
the kinetic energy $\hat{E}=\frac{1}{2}|\hat{\mathbf{v}}(\mathbf{y},t)|^{2}$
has explicit time dependence. As panels (b) and (c) of fig. \ref{fig:fig6}
show, the isosurfaces of the kinetic energy outline a vortical structure
parallel to the $z$-axis, the rotational axis of the $\mathbf{y}$-frame.
In contrast, the deformation kinetic energy preserves
the topology of its isosurfaces, regardless of which frame it is observed
in. Comparing the isosurfaces of the deformation kinetic energy in
the $\mathbf{x}$ and $\mathbf{y}$ frames, shown in panels (d)-(f)
and (g)-(i), the only difference is that for $t\neq0$, due to the
frame change, the whole computational domain is rotated around the
$z$-axis according to $\mathbf{y}=\mathbf{Q}^{T}(t)\mathbf{x}$.
Otherwise, the topologies of $E_{d}$ and $\hat{E}_{d}$
are identical in each case. 

\section{Conclusion}

We have derived a decomposition of an arbitrary velocity field $\mathbf{v}$
on a domain $U$ into its closest rigid body component, $\mathbf{v}_{RB}$,
and a deformation component, $\mathbf{v}_{d}=\mathbf{v}-\mathbf{v}_{RB}$.
The distance between $\mathbf{v}$ and $\mathbf{v}_{RB}$ is minimal
in the L$^{2}$ norm.

We have explicitly solved the optimization problem (\ref{eq:E-L equation})
for the angular velocity vector $\boldsymbol{\omega}$ of
the rigid body motion, which can equivalently be expressed with a
skew-symmetric tensor $\boldsymbol{\Omega}.$ This unique
rigid-body component renders the deformation component objective and
physically observable. Observability means that there is a distinguished
Euclidean observer who measures the velocity field to be equal to
its deformation component. 

We have illustrated the calculation of the deformation velocity on
examples. In particular, for planar flows, we have used unsteady,
polynomial solutions of the Navier--Stokes equation, a kinematic
model exhibiting material separation, and a satellite velocimetry
data set of the ocean surface velocity as examples. We have computed
the instantaneous deformation streamlines and compared the trajectories
of the deformation component to those of the original velocity field. 

With the help of the deformation velocity, we can also objectivize
most Eulerian quantities typically used in flow analysis just by computing
them on $\mathbf{v}_{d}$. This is equivalent to computing these
Eulerian quantities on the full velocity field $\mathbf{v}$ but in
a frame co-moving with $\mathbf{v}_{RB}$. We have illustrated the
objectivity of the deformation kinetic energy, deformation enstrophy
and deformation helicity for the classic ABC flow by recomputing these
quantities in a rotating frame. While the level set structure of the
kinetic energy changes substantially when observed from the rotating
frame, the deformation kinetic energy is indeed indifferent to this
observer change.

In almost all the examples considered here, the computation of the
deformation velocity field was simple and explicit. In some cases,
the deformation velocity provided more insight into the flow, in some
cases it did not. This is because truly unsteady, spatially complex
flows do not have a unique distinguished frame (\cite{lugt_dilemma_1979}).
We argue, however, that the variational problem we have posed and
solved here explicitly provides the closest possible rigid-body frame
in a well-defined physical sense. 

\section{Code availability}
The computations related to the implementation of the method on the presented examples can be found in the repository under the link
\newline \url{https://github.com/haller-group/DeformationVelocity}.
\section{Declaration of interests}
The authors report no conflict of interest.

\section{Acknowledgements}

The altimeter products were processed by SSALTO/DUACS and distributed
by AVISO+ (\href{https://www.aviso.altimetry.fr}{https://www.aviso.altimetry.fr}) with support from CNES.
We acknowledge support from the Turbulent Superstructures Program
(SPP1881) of the German National Science Foundation (DFG).

\section{Appendices\label{sec:Appendix}}

\subsection{Appendix A: The derivative of $L_{1}(\boldsymbol{\Omega},t)$\label{subsec:Appendix-A}}

To find the extremum of the functional $\widetilde{L}_{1}(\boldsymbol{\Omega},t)$
in eq. \ref{eq:L1_in_coordinates}, we first express it in terms of
the angular velocity $\boldsymbol{\omega}.$ The relationship 

\[
\boldsymbol{\Omega}(t)\mathbf{e}=\boldsymbol{\omega}(t)\times\mathbf{e},\quad\forall\mathbf{e}\in\mathbb{R}^{3},
\]
translates to coordinate components as
\begin{equation}
\boldsymbol{\Omega}_{ij}e_{j}=\varepsilon_{ijk}\omega_{j}e_{k},\label{eq:omega components}
\end{equation}
where $\varepsilon_{ijk}$ is the Levi--Civita symbol. Substituting
\ref{eq:omega components} into \ref{eq:L1_in_coordinates}, we obtain

\begin{align}
\widetilde{L}_{1}(\boldsymbol{\omega},t) & =\frac{1}{M}\int_{U}\Biggl\{-2(\mathbf{v}_{i}-\dot{\bar{\mathbf{x}}}_{i})\varepsilon_{ijk}\omega_{j}\left(\mathbf{x}_{k}-\bar{\mathbf{x}}_{k}\right)\label{eq:L1_components_omega}\\
 & +\varepsilon_{ijk}\omega_{j}\left(\mathbf{x}_{k}-\bar{\mathbf{x}_{k}}\right)\varepsilon_{ilm}\omega_{l}\left(\mathbf{x}_{m}-\bar{\mathbf{x}}_{m}\right)\Biggr\}\rho(\mathbf{x},t)dV.\nonumber 
\end{align}

We now seek the minimizer $\boldsymbol{\omega}(t)$ as the solution
to the system of three equations

\begin{align}
\frac{\partial}{\partial\omega_{n}}\widetilde{L_{1}}(\omega,t) =0,\qquad n=1,2,3.\label{eq:minimum equation}
\end{align}

We then differentiate the integrand in (\ref{eq:L1_components_omega}) to obtain

\begin{align}
\label{eq:derivative1}
 & \frac{\partial}{\partial\omega_{n}}\Biggl\{-2(\mathbf{v}_{i}-\dot{\bar{\mathbf{x}}}_{i})\varepsilon_{ijk}\omega_{j}\left(\mathbf{x}_{k}-\bar{\mathbf{x}}_{k}\right)+\varepsilon_{ijk}\omega_{j}\left(\mathbf{x}_{k}-\bar{\mathbf{x}}_{k}\right)\varepsilon_{ilm}\omega_{l}\left(\mathbf{x}_{m}-\bar{\mathbf{x}}_{m}\right)\Biggr\} \\
=&-2\varepsilon_{nki}(\mathbf{v}_{i}-\dot{\bar{\mathbf{x}}}_{i})(\mathbf{x}_{k}-\bar{\mathbf{x}}_{k})+2\omega_{n}(\mathbf{x}_{m}-\bar{\mathbf{x}}_{m})^{2} -2\omega_{k}(\mathbf{x}_{k}-\bar{\mathbf{x}}_{k})(\mathbf{x}_{n}-\bar{\mathbf{x}}_{n})\nonumber  .
\end{align}

Written in coordinate-invariant form, eq. (\ref{eq:derivative1})
becomes

\[
-2(\mathbf{x}-\bar{\mathbf{x}})\times(\mathbf{v}-\dot{\bar{\mathbf{x}}})+2\left(\left|\mathbf{x}-\bar{\mathbf{x}}\right|^{2}\right)\boldsymbol{\omega}-2(\mathbf{x}-\bar{\mathbf{x}})\otimes(\mathbf{x}-\bar{\mathbf{x}})\mathbf{\boldsymbol{\omega}}.
\]
Substituting this derivative of the integrand into (\ref{eq:minimum equation})
and dividing by $2$, we obtain eq. (\ref{eq:omega equation}), as
claimed.

\subsection{Appendix B: The second derivative of $\tilde{L}_{1}(\boldsymbol{\Omega},t)$\label{subsec:Appendix-B}}

To compute the Hessian of the function $\tilde{L}_{1}(\boldsymbol{\Omega},t)$,
we start from (\ref{eq:L1_components_omega}). Taking the derivative
of the integrand with respect to $\omega_{n},$ we have already obtained
the expression (\ref{eq:derivative1}). Further differentiation
with respect to $\omega_{p}$ then yields
\begin{align*}
 & \frac{\partial}{\partial\omega_{p}}\Biggl[2\varepsilon_{ink}(\mathbf{v}_{i}-\dot{\bar{\mathbf{x}}}_{i})(\mathbf{x}_{k}-\bar{\mathbf{x}}_{k})+2\omega_{n}(\mathbf{x}_{m}-\bar{\mathbf{x}}_{m})^{2}-2\omega_{k}(\mathbf{x}_{k}-\bar{\mathbf{x}}_{k})(\mathbf{x}_{n}-\bar{\mathbf{x}}_{n})\Biggr]\\
= & 2\delta_{np}(\mathbf{x}_{m}-\bar{\mathbf{x}}_{m})^{2}-2(\mathbf{x}_{p}-\bar{\mathbf{x}}_{p})(\mathbf{x}_{n}-\bar{\mathbf{x}}_{n}).
\end{align*}

Expressed in a coordinate-invariant form, we therefore obtain the
Hessian

\begin{align*}
\frac{\partial^{2}}{\partial\omega_{n}\partial\omega_{p}}\widetilde{L}_{1}(\boldsymbol{\omega},t) =2\delta_{np}(\mathbf{x}_{m}-\bar{\mathbf{x}}_{m})^{2}-2(\mathbf{x}_{p}-\bar{\mathbf{x}}_{p})(\mathbf{x}_{n}-\bar{\mathbf{x}}_{n})=\frac{2}{M}[\boldsymbol{\Theta}]_{np},
\end{align*}
as claimed.

\subsection{Appendix C: Transformation rule for the closest rigid-body angular
velocity vector\label{subsec:Appendix-C}}

Under a Euclidean frame change (\ref{eq:Frame change}), the velocity
$\mathbf{v}(\mathbf{x},t)$ and the average velocity $\bar{\mathbf{v}}(t)$
transform as 
\begin{equation}
\begin{aligned}\widehat{\mathbf{v}} & =\mathbf{Q}^{T}\left(\mathbf{v}-\dot{\mathbf{Q}}\mathbf{y}-\dot{\mathbf{b}}\right),\\
\widehat{\bar{\mathbf{v}}} & =\mathbf{Q}^{T}\left(\mathbf{\bar{v}}-\dot{\mathbf{Q}}\bar{\mathbf{y}}-\dot{\mathbf{b}}\right).
\end{aligned}
\label{eq:v-transform}
\end{equation}
Since $\mathbf{Q}^{T}\dot{\mathbf{Q}}$ is skew-symmetric, it has
a dual vector associated to it, $\dot{\mathbf{q}}$, defined as

\begin{equation}
\dot{\mathbf{Q}}\mathbf{Q}^{T}\mathbf{e}=\dot{\mathbf{q}}\times\mathbf{e}\qquad\forall\mathbf{e}\in\mathbb{R}^{3}.\label{eq:qdot_def}
\end{equation}
Substituting the transformation rules (\ref{eq:transf of moment of inertia tensor}),
(\ref{eq:v-transform}) into (\ref{eq:omega equation}),
we obtain
\begin{align}
\hat{\boldsymbol{\Theta}}\hat{\boldsymbol{\omega}} & =M\overline{(\mathbf{y}-\bar{\mathbf{y}})\times(\hat{\mathbf{v}}-\hat{\bar{\mathbf{v}}})}\nonumber \\
 & =M\overline{\mathbf{Q}^{T}(\mathbf{x}-\overline{\mathbf{x}})\times\mathbf{Q}^{T}\left(\mathbf{v}-\overline{\mathbf{v}}-\dot{\mathbf{Q}}\mathbf{Q}^{T}(\mathbf{x}-\overline{\mathbf{x}})\right)}.\label{eq:omega-transf1}
\end{align}
We now recall that for any rotation matrix $\mathbf{Q}$ and for arbitrary
vectors $\mathbf{a}$ and $\mathbf{b}$, we have $\mathbf{Qa}\times\mathbf{Qb=\mathbf{Q(a\times b)}}.$
Furthermore, for any three vectors $\mathbf{a},\mathbf{b},\mathbf{c}\in\mathbb{R}^{3}$,
we have $\mathbf{a\times(b\times c)=\mathbf{b(a\cdot c)}-}\mathbf{c}(\mathbf{a\cdot b}).$
With these identities, we can rewrite eq. (\ref{eq:omega-transf1})
as 
\begin{align*}
\hat{\boldsymbol{\Theta}}\hat{\boldsymbol{\omega}} & =M\mathbf{Q}^{T}\overline{(\mathbf{x}-\overline{\mathbf{x}})\times\left(\mathbf{v}-\overline{\mathbf{v}}-\dot{\mathbf{Q}}\mathbf{Q}^{T}(\mathbf{x}-\overline{\mathbf{x}})\right)}\\
 & =M\mathbf{Q}^{T}\overline{\left(\mathbf{x}-\overline{\mathbf{x}}\right)\times\left(\mathbf{v}-\overline{\mathbf{v}}\right)-\dot{\mathbf{q}}\left|\mathbf{x}-\overline{\mathbf{x}}\right|^{2}+\left(\mathbf{x}-\overline{\mathbf{x}}\right)\left[\mathbf{\dot{q}\cdot}\left(\mathbf{x}-\overline{\mathbf{x}}\right)\right]}\\
 & =M\mathbf{Q}^{T}\overline{(\mathbf{x}-\bar{\mathbf{x}})\times(\mathbf{v}-\overline{\mathbf{v}})}-\mathbf{Q}^{T}\boldsymbol{\Theta}\mathbf{\dot{q}}\\
 & =M\mathbf{Q}^{T}\left[\frac{1}{M}\boldsymbol{\Theta}\mathbf{\boldsymbol{\omega}}-\frac{1}{M}\boldsymbol{\Theta}\mathbf{\dot{q}}\right].\\
\mathbf{Q}^{T}\boldsymbol{\Theta}\mathbf{Q}\hat{\boldsymbol{\omega}} & =\mathbf{Q}^{T}\boldsymbol{\Theta}\mathbf{\boldsymbol{\omega}}-\mathbf{Q}^{T}\boldsymbol{\Theta}\mathbf{\dot{q}}
\end{align*}
 Substituting the transformation formula (\ref{eq:transf of moment of inertia tensor})
for the tensor $\hat{\boldsymbol{\Theta}}$ into the left-hand
side of this last equation then yields the transformation formula
\[
\hat{\boldsymbol{\omega}}=\mathbf{\mathbf{Q}}^{T}\left(\boldsymbol{\omega}-\dot{\mathbf{q}}\right)
\]
 for the angular velocity $\boldsymbol{\omega}$. 

\subsection{Appendix D: Transformation rule of the deformation velocity\label{subsec:Appendix-D}}

Under a Euclidean frame change (\ref{eq:Frame change}), the deformation
velocity field $\mathbf{v}_{\text{\ensuremath{d}}}(\mathbf{x},t)$
defined in (\ref{eq:definition of deformation velocity}) can be computed
in the $\mathbf{y}$-frame as

\begin{align}
\hat{\mathbf{v}}_{d} & =\mathbf{\hat{v}}-\hat{\bar{\mathbf{v}}}-\mathbf{\boldsymbol{\widehat{\omega}}}\times(\mathbf{y}-\mathbf{\overline{y}})\nonumber \\
 & =\mathbf{Q}^{T}\left[\mathbf{v}-\overline{\mathbf{v}}-\dot{\mathbf{Q}}\mathbf{Q}^{T}(\mathbf{x}-\overline{\mathbf{x}})\right]-\mathbf{\mathbf{Q}}^{T}\left(\boldsymbol{\omega}-\dot{\mathbf{q}}\right)\times\mathbf{Q}^{T}(\mathbf{x}-\overline{\mathbf{x}})\nonumber \\
 & =\mathbf{Q}^{T}\left[\mathbf{v}-\overline{\mathbf{v}}-\boldsymbol{\omega}\times(\mathbf{x-\overline{x}})\right]\nonumber \\
 & =\mathbf{Q}^{T}\mathbf{v}_{\text{\ensuremath{d}}},\label{eq:Def velocity is obj}
\end{align}
 which proves the objectivity of $\mathbf{v}_{\text{\ensuremath{d}}}(\mathbf{x},t)$. 

\subsection{Appendix E: The full velocity field in a frame co-moving with $\mathbf{v}_{RB}$
\label{subsec:Appendix-E}}

Under the observer change defined in (\ref{eq:change to special observer})-(\ref{eq:specific frame change}),
the transformed velocity field becomes 
\begin{align*}
\hat{\mathbf{v}} & =\mathbf{Q}_{RB}^{T}\left(\mathbf{v}-\dot{\mathbf{Q}}_{RB}\mathbf{y}-\dot{\mathbf{b}}_{RB}\right)\\
 & =\mathbf{Q}_{RB}^{T}\left(\mathbf{v}-\dot{\mathbf{Q}}_{RB}\mathbf{Q}_{RB}^{T}\left(\mathbf{x}-\mathbf{b}_{RB}\right)-\dot{\mathbf{b}}_{RB}\right)\\
 & =\mathbf{Q}_{RB}^{T}\left(\mathbf{v}-\boldsymbol{\omega}\times\mathbf{x}+\boldsymbol{\omega}\times\bar{\mathbf{x}}-\mathbf{\bar{v}}\right)=\mathbf{Q}_{RB}^{T}\textbf{\ensuremath{\mathbf{v}_{d}}}=\hat{\mathbf{v}}_{d},
\end{align*}
 where we have used formula (\ref{eq:Def velocity is obj}).

\subsection{Appendix F: Deformation velocity of the deformation velocity \label{subsec:Appendix-F}}

First, as a special case, we formally calculate $\boldsymbol{\omega}$
for a rigid-body velocity field 
\begin{equation}
\mathbf{v}(\mathbf{x},t)=\mathbf{p}\times(\mathbf{x}-\bar{\mathbf{x}})+\bar{\mathbf{v}}.\label{eq:linear velocity}
\end{equation}
Noting that 

\begin{align*}
(\mathbf{x}-\bar{\mathbf{x}})\times(\mathbf{v}-\bar{\mathbf{x}}) & =\mathbf{(\mathbf{x}-\bar{\mathbf{x}})\times p}\times(\mathbf{x}-\bar{\mathbf{x}})\\
 & =\left|\mathbf{x}-\bar{\mathbf{x}}\right|^{2}\mathbf{p}+(\mathbf{x}-\bar{\mathbf{x}})\otimes(\mathbf{x}-\bar{\mathbf{x}})\mathbf{p},\\
\end{align*}
we find from (\ref{eq:optimal solution-1}) that
\begin{align}
\boldsymbol{\omega} & =M\boldsymbol{\Theta}^{-1}\left[\overline{\left|\mathbf{x}-\bar{\mathbf{x}}\right|^{2}\mathbf{p}+(\mathbf{x}-\bar{\mathbf{x}})\otimes(\mathbf{x}-\bar{\mathbf{x}})\mathbf{p}}\right]\label{eq:rigidbody omega}\\
 & =M\boldsymbol{\Theta}^{-1}\frac{1}{M}\boldsymbol{\Theta}\mathbf{p}=\mathbf{p},\nonumber 
\end{align}
i.e., $\boldsymbol{\omega}$ coincides with the angular velocity
of the rigid body. 

Therefore, noting that (\ref{eq:optimal solution-1}) shows $\boldsymbol{\omega}$
to be a homogeneous linear function of the velocity field $\mathbf{v}$,
we obtain
\[
\boldsymbol{\omega}^{(2)}=\boldsymbol{\omega}-\boldsymbol{\omega}=\mathbf{0}.
\]
 In other words, the deformation velocity has a vanishing closest
rigid body component under our optimization principle, given that
we also have $\mathbf{v}(\bar{\mathbf{x}},t)=\bar{\mathbf{v}}$ in
eq. (\ref{eq:linear velocity}).

{ 
\subsection{Appendix G: Generalizing the objective function}
As a straightforward generalization to the mass-based $L^2$ norm \eqref{eq:sobolevnorm}, we consider here the following family of Sobolev norms to measure the closeness of two velocity fields:
\begin{align}
    \left\Vert \mathbf{f}\right\Vert _{H^{k}}^{2}&=\frac{1}{M}\int_{U}\left|\mathbf{f}(\mathbf{x},t)\right|^{2}dm+\alpha\frac{1}{M}\int_{U}\left|\mathbf{D}_{\mathbf{x}}\mathbf{f}(\mathbf{x},t)\right|^{2}dm \\ &+\frac{1}{M}\sum_{i=2}^{k}\beta_{i}\int_{U}\left|\mathbf{D}_{\mathbf{x}}^{k}\mathbf{f}(\mathbf{x},t)\right|^{2}dm.
    \label{eq:alphanorm}
\end{align}
This $H^k$ norm is defined for integrable functions over U whose derivatives up to order $k$ are also integrable. The constants $\alpha,\beta_{i}\geq0$ ensure the same physical dimension for all terms in $\left\Vert \mathbf{f}\right\Vert _{H^{k}}^{2}$ and also serve to assign different weights to the different spatial scales in $\mathbf{f}(\mathbf{x},t)$. Specifically, $\alpha=\beta_{1}=\ldots=\beta_{k}=0$ eliminates the consideration of smaller scales in the norm of $\mathbf{f}(\mathbf{x},t)$, whereas $\alpha,\beta_{i}\to\infty$ gives full weight to the smaller scales. 

Therefore, the objective function to be minimized becomes 
\begin{equation}
    L_{H^k}(\boldsymbol{\Omega}, \mathbf{x}_A,t) = \left \Vert \mathbf{v}(\mathbf{x},t) - \mathbf{v}_{RB}(\mathbf{x},t) \right \Vert_{H^{k}}^{2}.
\end{equation}
Due to the spatial linearity of $\mathbf{v}_{RB}$, for any $k\geq1$, extrema of $L_{H^k}(\boldsymbol{\Omega}(t),t)$ are defined by the equation
\begin{equation}
    \frac{\partial}{\partial\boldsymbol{\Omega}}L_{H^k}(\boldsymbol{\Omega},t)\equiv\frac{\partial}{\partial\boldsymbol{\Omega}}L_{H^1}(\boldsymbol{\Omega},t)=0.
\end{equation}
Therefore, the weights given to the higher-order derivatives in \eqref{eq:alphanorm} can be set as $\beta_1 =  ... = \beta_k = 0$ without loss of generality. Therefore, the only remaining parameter to determine is the weight of the derivative term, $\alpha$. 

With a simple extension of the derivation outlined in Section \ref{sec:Derivation-of-the}, the closest rigid-body angular velocity with respect to the $H^1$ norm is 
\begin{equation}
    \boldsymbol{\omega}_\alpha =  M\boldsymbol{\Theta_\alpha}^{-1}\overline{(\mathbf{x}-\bar{\mathbf{x}})\times(\mathbf{v}-\bar{\mathbf{v}}) + \alpha \nabla \times \mathbf{v}},
    \label{eq:omegaalpha}
    \end{equation}
    where the moment of inertia tensor $\boldsymbol{\Theta}_0$ is modified to $\boldsymbol{\Theta}_\alpha = \boldsymbol{\Theta}_0 + 2\alpha \mathbf{I}$. 
    
    In Fig. \ref{fig:fig7}, we compare the optimal rigid-body angular velocity for various values of $\alpha$. We take the AVISO dataset analyzed in Fig. \ref{fig:fig4} and compute $\omega$ again. For both of the domains considered, we see that $\omega_\alpha$ does not change significantly with $\alpha$. Therefore, the basic conclusions we drew from the $\alpha=0$ result of Fig. \ref{fig:fig4} extends to any $\alpha$. Optimizing over the whole domain gives small negative values of the angular velocity, while optimizing over the neighborhood of the mesoscale eddy gives a larger positive angular velocity. In light of this, the resulting deformation velocities are also similar to each other for non-zero values of $\alpha$. In the main body of this paper, we have decided to set $\alpha=0$ in the main results to remain in line with our physical motivation. 
\begin{figure}[h!]
    \centering
    \includegraphics[width = 0.6 \textwidth]{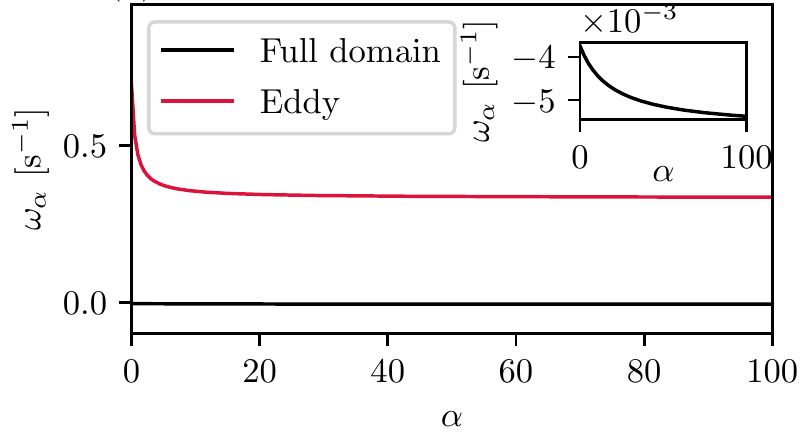}
    \caption{{The rigid-body angular velocity defined by \eqref{eq:omegaalpha} calculated for the AVISO dataset featured in Fig. \ref{fig:fig4}. The black curve shows the angular velocity calculated over the full domain shown in Fig. \ref{fig:fig4}. The red curve shows the result of the calculation restricted to the neighborhood of the mesoscale eddy shown in the insets of Fig. \ref{fig:fig4}.}}
    \label{fig:fig7}
\end{figure}
    
}


\end{document}